# Machine Learning in Official Statistics

Martin Beck[1], Florian Dumpert[2], Joerg Feuerhake[3]

As of 13 December 2018.

## Abstract

In the first half of 2018, the Federal Statistical Office of Germany (Destatis) carried out a "Proof of Concept Machine Learning" as part of its Digital Agenda. A major component of this was surveys on the use of machine learning methods in official statistics, which were conducted at selected national and international statistical institutions and among the divisions of Destatis. It was of particular interest to find out in which statistical areas and for which tasks machine learning is used and which methods are applied. This paper is intended to make the results of the surveys publicly accessible.

[1] Federal Statistical Office of Germany; corresponding author martin.beck@destatis.de
[2] Department of Mathematics, University of Bayreuth, Germany
[3] Federal Statistical Office of Germany

# 1 Introduction

## 1.1 Digital Agenda of the Federal Statistical Office of Germany

On 10 October 2017, the development of a Digital Agenda of the Federal Statistical Office of Germany (Destatis) has started (Statistisches Bundesamt 2018). One of many topics that were intensively discussed was Machine Learning. In a meeting at 13–15 November 2017, the office and department heads of Destatis evaluated and prioritised 59 measures of the Digital Agenda according to their benefits and costs. A "Proof of Concept Machine Learning" was given high priority and classified as one of four lighthouse projects of the Digital Agenda. The content specification was "Proof of Concept Machine Learning – Set up Proof of Concept for Machine Learning, e.g. in business statistics, to perform automatic categorization and improve analysis potential". The deadline for completion of the project was set for mid-2018.[4]

## 1.2 Proof of Concept Machine Learning

The topics "Machine Learning" and even more "Artificial Intelligence" are very comprehensive and can hardly be overlooked. To ensure that the project remained manageable and could be completed within the available time and with the capacities at hand, it was first necessary to narrow down and operationalise the task. The project team agreed to concentrate exclusively on machine learning and to focus on its applicability in official statistics, the latter, however, not limited to business statistics. Finally, the task "Proof of Concept Machine Learning" was interpreted as follows: Verification of the applicability of machine learning in the processes of official statistics. After completion of the proof of concept, an overview of the potential applications in official statistics should be available.

At the beginning of the project, the Federal Statistical Office of Germany already had concrete but limited experience with the use of machine learning in business register, crafts statistic, and the survey of earnings structure (Dumpert and Beck 2017). However, in order to be able to investigate their possible uses across the board, the statistical experts of Destatis should first be provided with information about machine learning methods themselves and the potential fields of application. One crucial input were the results of an inventory of the application of machine learning methods in national and international statistical institutions. This knowledge base was then used to conduct an in-house survey on the potential use of machine learning methods. The method and the complete results of the inventory and the in-house survey are described in the remaining sections.

---

[4] The "Proof of Concept Machine Learning" was completed by 31 July 2018. The final report (Beck, Dumpert, and Feuerhake 2018) was submitted to and accepted by the president of Destatis.

## 2 Survey on the use of machine learning in statistical institutions

### 2.1 Method

In order to get a picture of use cases for machine learning currently pursued in national and international statistics producing institutions, a survey was conducted. The aim was to obtain information on relevant fields of application and on methods used, and thereby to learn for further action in the Federal Statistical Office. On the national level, the 14 Statistical Offices of the Länder were contacted on 13 March 2018 by means of a structured questionnaire (Excel file). Furthermore 18 additional statistical producers (mainly so-called ONAs – Other National Authorities) received the same questionnaire. The addressees were informed that the Federal Statistical Office carries out a "Proof of Concept Machine Learning" and, in this context, conducts surveys of national and international statistical institutions in order to assess the possible uses of machine learning in official statistics. The surveyed institutions were asked to provide the following information for each of their machine learning related projects:

| Institution: | Name or abbreviation of the statistics-producing organization. |
| --- | --- |
| Project name: | Name of the machine learning project. |
| Description: | Short description of the machine learning project (e.g. goal, approach, methods). |
| Application: | What is to be achieved with the machine learning method, e.g. classification, regression, clustering, etc.? |
| Status: | Status of the project<br>- Productive<br>- Experiment<br>- Test<br>- Idea |
| Method: | Which machine learning method was used, e.g.:<br>- Support vector machine (SVM)<br>- Decision trees<br>- Random forest<br>- Neural networks |
| Software: | Which software is used (e.g. R, Python, etc.)? |
| Source (Link): | Are there any publications or descriptions available? |

All institutions responded to the request, see Table 1.

| Institution | Application of machine learning | |
|---|---|---|
| | yes | no |
| Federal Employment Agency (BA) | | x |
| Federal Office for Migration and Refugees (BAMF) | x | |
| Federal Office of Consumer Protection and Food Safety (BVL) | | x |
| Federal Office for Economic Affairs and Export Control (BAFA) | | x |
| Federal Office for Agriculture and Food (BLE) | | x |
| German Central Bank (Deutsche Bundesbank) | x | |
| FDZ German Pension Insurance | | x |
| FDZ Institute for the Future of Work (FDZ IZA) | | x |
| FDZ SOEP | | x |
| GESIS | x | |
| Institute for Employment Research (IAB) | x | |
| Julius Kühn Institute | | x |
| Federal Motor Transport Authority | | x |
| Robert Koch Institute | x | |
| Stifterverband|Wissenschaftstatistik | | x |
| Thünen Institute | | x |
| German Environment Agency | | x |
| Centre for European Economic Research (ZEW) | x | |
| Hessian Statistical Office | x | |
| Remaining 13 statistical offices of the Länder | | x |

Table 1: Survey of national statistical institutions

In the same way, the statistical offices of the 27 EU member states, the four EFTA countries, six selected non-European countries (Australia, Canada, Israel, Japan, New Zealand and USA) as well as Eurostat and the OECD were contacted. The respondents answered somewhat reluctantly. Despite three reminders, six[5] offices unfortunately did not take part in the survey. 21 reported machinelearning projects and 12 reported no activity in the field (see Table 2).

Another source of data was the "Machine Learning Documentation Initiative" (Chu and Poirier 2015), which is currently being continued by Valentin Todorov (UNIDO) who thankfully provided us an updated but yet unpublished version in January 2018. In addition, an attempt was made to obtain information about research on the websites of statistical offices that did not respond to the survey and are not listed in the "Machine Learning Documentation Initiative". Despite all efforts, it was not possible to obtain data for Bulgaria, Greece and Malta. Nevertheless, the information gathered provides a broad picture of the international use of machine learning methods in official statistics.

---

[5] France (INSEE) reported five machine learning applications after the final report of the "Proof of Concept Machine Learning" was completed. The Netherlands (CBS) added detailed information about their machine learning projects in early December 2018. Both replies are included in this overview but were not part of the final report of the "Proof of Concept Machine Learning". Therefore the detailed results differ slightly. The overall results and conclusion are nevertheless unaffected.

| Country | Application of machine learning | | No feedback |
|---|---|---|---|
| | yes | no | |
| Australia | x | | |
| Austria | x | | |
| Belgium | x | | |
| Bulgaria | | | x |
| Canada | x | | |
| Croatia | | x | |
| Cyprus | | x | |
| Czechia | | x | |
| Denmark | x | | |
| Estonia | | x | |
| Eurostat | | | x |
| Finland | x | | |
| France | x | | |
| Greece | | | x |
| Hungary | | x | |
| Iceland | x | | |
| Ireland | | x | |
| Israel | | x | |
| Italy | | | x |
| Japan | x | | |
| Latvia | x | | |
| Liechtenstein | | x | |
| Lithuania | | x | |
| Luxembourg | x | | |
| Malta | | | x |
| Netherlands | x | | |
| New Zealand | x | | |
| Norway | x | | |
| OECD | x | | |
| Poland | | x | |
| Portugal | x | | |
| Romania | x | | |
| Slovakia | | x | |
| Slovenia | | x | |
| Spain | x | | |
| Sweden | x | | |
| Switzerland | x | | |
| UK | | | x |
| USA | x | | |

Table 2: Survey of international statistical institutions

## 2.2   Results of the survey of the statistical offices of the Länder and other national institutions

All 14 statistical offices of the Länder were contacted within the scope of the survey. 13 reported no projects in the field of machine learning. The Hessian Statistical Office is the only one to test a procedure. Here, data on enterprises in the business register obtained by means of web scraping is evaluated. Methods of machine learning are used here, among other things, to carry out cross-statistical coherence checks and generate new features.

As mentioned above, we inquired information on the use of machine learning from several ONAs. We received positive feedback from the Federal Office for Migration and Refugees (BAMF), the German Central Bank (Deutsche Bundesbank)[6], the GESIS – Leibniz Institute for Social Sciences, the Institute for Employment Research (IAB), the Robert Koch Institute (RKI) and the Centre for European Economic Research (ZEW). They reported a total of 36 projects using machine learning techniques. Five of these applications are in productive operation. These are procedures for imputing missing information on working time, for predicting the duration of unemployment and for classifying labour market regions, which are carried out at the Institute for Employment Research (IAB). The Robert Koch Institute uses machine learning productively for the automatic detection of outbreaks of infectious diseases and for the analysis of molecular data.

A further 21 projects are ongoing research projects in the narrower sense. The remaining ten projects are feasibility studies or the development of prototypes with potential for productive operation.

---

[6] The Deutsche Bundesbank provided information on five projects, which were included in the aggregated results presented here. Due to an agreement with the Deutsche Bundesbank, the detailed results in tabular form in section 4.1 only include those projects for which publications are already available.

| Institute | Number of applications |
|---|---|
| GESIS | 16 |
| IAB | 8 |
| German Central Bank | 5 |
| RKI | 4 |
| ZEW | 2 |
| BAMF | 1 |
| Total | 36 |

Table 3: Machine learning applications by institution

| Used machine learning methods (multiple answers possible) | Number |
|---|---|
| Decision tree based methods[7] | 13 |
| Random forest | 13 |
| Neural networks | 11 |
| SVM | 10 |
| Other | 12 |
| Total | 59 |

Table 4: Machine learning methods used

| Type of application (multiple answers possible) | Number |
|---|---|
| Classification | 19 |
| Identification | 11 |
| Clustering | 6 |
| Text analysis | 9 |
| Regression | 4 |
| Other | 7 |
| Total | 56 |

Table 5: Type of application

According to the responses, machine learning methods were most frequently used to identify or classify units. Another important field of application is regression. In addition to random forests, other decision tree based methods such as gradient boosting are most frequently used. Neural networks and support vector machines are also often used.

Although the participating institutions pursue very different tasks and therefore have very different project objectives, it can be stated that problems tackled with machine learning procedures are often similar ones. The methods used are also often similar.

The detailed results are presented in Section 4.1.

## 2.3   Results of international institutions

The majority of the international statistics producing institutions inquired were national statistics authorities. It turned out that these statistical offices carry out projects with machine learning methods to varying degrees. Statistics Canada reported by far the largest number of such projects. Statistics Netherlands and the U.S. Bureau of Labour Statistics also run a relatively large number of projects (see Table 6).

---

[7] The number 13 for decision tree based methods and random forest is randomly identical. The decision tree based methods are always explicitly not random forest.

| Institution | Number of projects |
|---|---|
| Statistics Canada | 36 |
| Statistics Netherlands | 16 |
| U.S. Bureau of Labor Statistics | 11 |
| Stats NZ | 9 |
| U.S. Department of Agriculture NASS | 7 |
| Australian Bureau of Statistics | 6 |
| Federal Statistical Office of Switzerland | 6 |
| INSEE (France) | 5 |
| National Institute of Statistics Romania | 4 |
| Statistics Austria | 4 |
| Statistics Portugal | 4 |
| Statistics Spain (INE) | 3 |
| Statistics Sweden | 3 |
| Eurostat | 2 |
| STATEC (Luxembourg) | 2 |
| Statistics Finland | 2 |
| Statistics Iceland | 2 |
| Statistics Poland | 2 |
| Bureau of Economic Analysis (USA) | 1 |
| Central Statistical Bureau of Latvia | 1 |
| Central Statistics Office of Ireland | 1 |
| Hungarian Central Statistical Office | 1 |
| Italian National Institute of Statistics | 1 |
| National Statistics Center, Japan | 1 |
| OECD | 1 |
| ONS (UK) | 1 |
| Statistics Belgium | 1 |
| Statistics Denmark | 1 |
| Statistics Norway | 1 |
| U.S. Census Bureau | 1 |
| Total | 136 |

Table 6: Survey and research of international statistical organisations, number of projects by institution

It is of particular interest which problems are to be solved with machine learning methods and the statistical areas in which these are used. It turned out that the majority of the projects are cross-statistical, i.e. they provide or are to provide procedures that can be used for several statistics. Frequently mentioned statistical areas are household and business statistics.

Machine learning methods are often used for classification, identification and imputation. The identification of units is often mentioned in connection with microdata linking. The productive status is also relevant for the evaluation of individual projects. 21 of the projects mentioned are already in productive use, another 28 are in development for productive use. 61 projects are in the experimental stage and another 26 are currently formulated as ideas.

| Statistics | Number of applications |
|---|---|
| Cross-statistical | 26 |
| Labour market | 15 |
| Household statistics | 14 |
| Agricultural statistics | 10 |
| Business statistics | 15 |
| Census | 8 |
| Branch classification | 7 |
| Price statistics | 5 |
| Traffic statistics | 4 |
| Other | 32 |
| Total | 136 |

Table 7: Machine learning applications according to subject statistics

| Type of application (multiple answers possible) | Number |
|---|---|
| Classification | 78 |
| Imputation | 22 |
| Microdata linking | 15 |
| Clustering | 9 |
| Text analysis | 8 |
| Regression | 6 |
| Identification | 4 |
| Dimension reduction | 2 |
| Other | 17 |
| Total | 161 |

Table 8: Type of application

| Project status | Number of applications |
|---|---|
| Idea | 26 |
| Experiment | 61 |
| In development | 28 |
| Productive | 21 |
| Total | 136 |

Table 9: Number of applications by project status

| Used machine learning methods (multiple answers possible) | Number |
|---|---|
| Random forest | 37 |
| Neural networks | 22 |
| SVM | 22 |
| Decision tree methods | 20 |
| Nearest-neighbour approaches | 12 |
| Bayesian approaches | 6 |
| Natural language processing | 5 |
| Cluster method | 2 |
| Other | 45 |
| Total | 171 |

Table 10: Machine learning methods used

Among the machine learning methods most frequently mentioned are random forests, methods that use neural networks, support vector machines and other decision tree based methods.

In summary, it can be stated that machine learning is used or tested in many statistical areas in international statistical offices. Usually, classification, identification or imputation tasks are tackled. Often decision tree based methods, neural networks or support vector machines are used.

The detailed results are presented in Section 4.2. The replies of the international statistical offices have been adopted unchanged by the authors of this paper and have virtually not been edited.

## 2.4 Assignation to the phases of the GSBPM

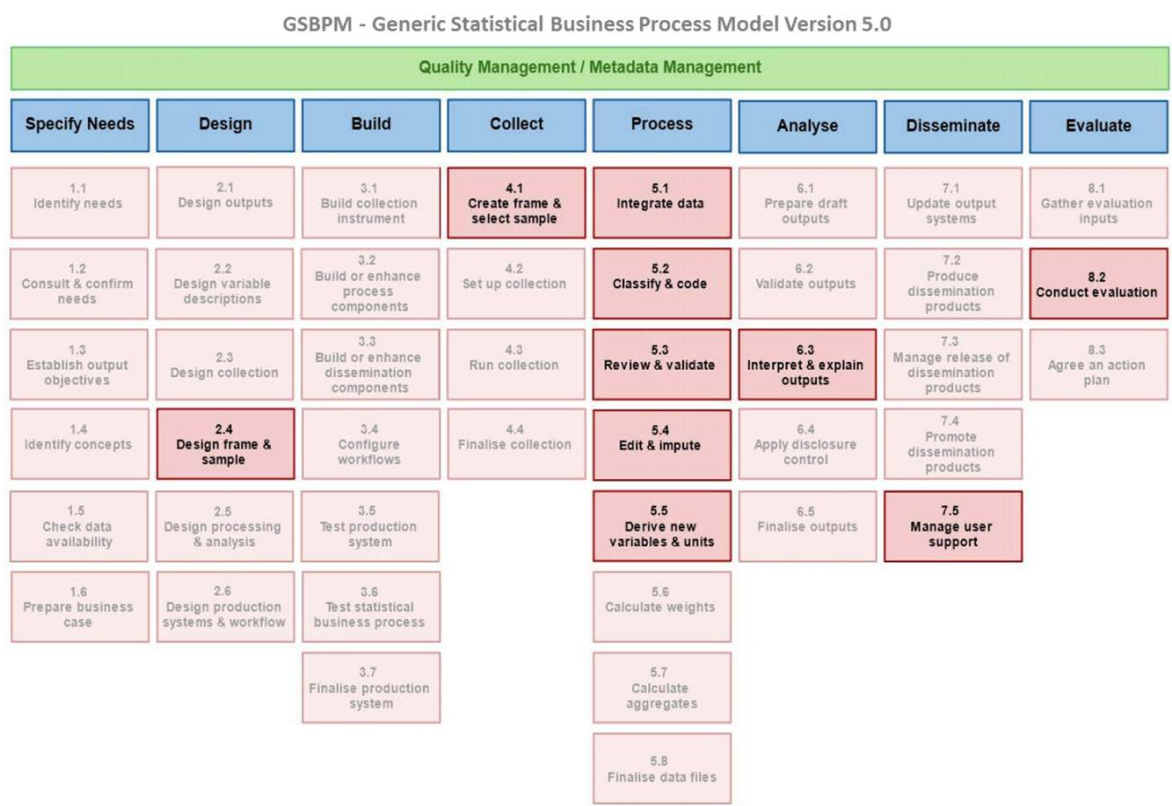

Figure 1: Sub processes of the GSBPM where machine learning projects have been reported by the responding national and international institutions

Since the national and international institutions are involved in producing statistics, the reported projects can be assigned to the phases of the GSBPM. Figure 1 shows the sub-processes of the GSBPM that are supported by these projects. It can be seen that machine learning procedures are mainly used for data acquisition, data preparation and result analysis. Further projects tackled tasks in the statistical conception, the organisation of the user service and in the evaluation. In connection with the GSBPM, Statistics Canada's paper "Machine-Learning in Surveys Steps" lists additional use cases in which machine learning methods can be used (Statistics Canada 2018).

The following conclusions can be drawn from the queries at the statistical offices of the Länder, the national and international statistics producing institutions: projects that automatically classify features, that impute values or identify units using random forests or similar tree based methods, support vector machines or neural networks are currently widespread. Use cases can be found in almost all subject statistics. As a rule, processes in statistical conception, data acquisition and analysis as well as statistical dissemination and evaluation are supported by machine learning procedures.

# 3 In-house survey

## 3.1 Method

In order to identify possible areas of application for machine learning methods in the various statistics areas an in-house survey was conducted among the divisions of the Federal Statistical Office of Germany. On 9 May 2018, all 29 divisions were contacted for the query, who also replied.

## 3.2 Outcomes

From the divisions surveyed, 16 reported no relevant projects, usually with the indication that no suitable applications are currently apparent. Five times a lack of expertise or a lack of time either to build up expertise or to implement projects was also cited as a reason. A lack of information about machine learning in general, however, was hardly mentioned as a reason. All in all, even divisions that did not report any projects or project ideas are considered to have potential for corresponding applications.

The remaining 13 divisions reported 31 possible applications. 25 of which were ideas on how to use machine learning. Six projects are already in an experimental or test phase. Since many proposals are first ideas, no final statement can be made about the methods used. Almost all divisions that reported project ideas also indicated that they would need external expertise in the possible implementation.

The project ideas often aim at the automatic classification of features or the identification of units (duplicates, outliers). These are also the applications very often mentioned in the feedback from the national and international statistics producing institutions.

In summary, it can be said that there are many promising approaches and ideas for the use of machine learning in the Federal Statistical Office of Germany. At the moment, however, there seems to be a bottleneck in the development or provision of expertise in this field.

The detailed results of the in-house survey are presented as part of Section 4.1. We have only included projects that have already reached a certain degree of maturity.

# 4 Feedback from national and international query

## 4.1 Application of machine learning in national statistical institutions

| Institution | Project name | Description of the project | Application | Status | Method | Software |
|---|---|---|---|---|---|---|
| Federal Statistical Office of Germany | Assignment of enterprises in the business register to the institutional sectors[8] | Support vector machines are used for the assignment of enterprises to the institutional sectors according to the European System of National Accounts (ESA) 2010. In this way, official statistics in Germany are entering new territory and creating an opportunity to replace costly searches carried out by humans to classify many individual cases in the future through this method of machine learning. | Binary classification | Productive | SVM | R |
| Federal Statistical Office of Germany | Recognition of irrelevant enterprises in the craft statistics[9] | The craft statistics are currently determined completely from administrative data. However, not all enterprises included in deliveries from chambers of crafts are relevant for craft statistics. Irrelevant enterprises have to be identified by the statistical offices of the Länder. Since the work on manual classification ties up personnel resources to a considerable extent, the question arose as to whether enterprises can be classified automatically with sufficient accuracy. Support Vector Machines in conjunction with Random Forests are currently used to classify enterprises as relevant or irrelevant with relatively little personnel effort. The training and test data originates from the statistical business register. After a few steps to reduce the dimensions, a Support Vector Machine is adapted to a data set of approximately 80,000 units with 30 characteristics. | Binary classification | Productive | SVM, Random Forest | R |

---

[8] Dumpert, F., von Eschwege, K., Beck, M. (2016). Einsatz von Support Vector Machines bei der Sektorzuordnung von Unternehmen. WISTA Nr. 1/2016, 87–97, https://www.destatis.de/DE/Publikationen/WirtschaftStatistik/2016/01/EinsatzSupportVectorMachines_012016.pdf (last accessed 10 December 2018).
[9] Feuerhake, J., Dumpert, F. (2016). Erkennung nichtrelevanter Unternehmen in den Handwerksstatistiken. WISTA Nr. 2/2016, 79–94, https://www.destatis.de/DE/Publikationen/WirtschaftStatistik/2016/02/NichtrelevanteUnternehmen_Handwerk_022016.pdf (last accessed 10 December 2018).

| Institution | Project name | Description of the project | Application | Status | Method | Software |
|---|---|---|---|---|---|---|
| Federal Statistical Office of Germany | Estimation of the interruption of employment due to motherhood[10] | In order to approach the missing characteristic "true work experience of a female employee" for the estimation of the adjusted gender pay gap, the data of the Structure of Earnings Survey (SES) should be enriched by a woman-specific variable "employment interruption due to the birth of a child (yes/no)". This would enable the Federal Statistical Office to take this form of career interruption into account when estimating the adjusted gender pay gap.<br>The 2012 microcensus contains the (albeit voluntary) characteristic "Did you give birth to children?", which was interpreted as an indication of the existence of a career interruption. The microcensus as external source therefore provided the training and test data to classify female employees as mothers and non-mothers. The experimental estimation of maternity among female employees included in the SES sample for 2014 yielded results that were included in the model for calculating the adjusted gender pay gap, but only reduced it in the range of the statistical uncertainty that existed anyway. | Binary classification | Research paper | SVM, Random Forest, logistic regression | R |
| Federal Statistical Office of Germany | Estimation of citizenship (German/foreigner) of employees in the Structure of Earnings Survey[11] | Another characteristic not currently covered by the Structure of Earnings Survey (SES) is the nationality of employees. Again, the microcensus as external source provided the training and test data to classify employees as Germans and foreigners. The following difficulty arose: The group of respondents in the microcensus with purely non-German citizenship is so small that each classification procedure "sacrifices" this group to achieve a low rate of misclassification; in other words, all foreigners are classified as Germans. Different approaches to dealing with so-called unbalanced data were therefore examined with limited access. However, the results were not convincing: Obviously, foreigners and Germans are too "similar" in the microcensus to be classified sufficiently well with the classical methods and machine-learning procedures used so far on the basis of the common characteristics of SES and microcensus. Studies based on other data sources have not yet been carried out. | Binary classification | Experiment | SVM (among others) | R |

---

[10] Finke, C., Dumpert, F., Beck, M. (2017). Verdienstunterschiede zwischen Männern und Frauen. WISTA Nr. 2/2017, 43–61, https://www.destatis.de/DE/Publikationen/WirtschaftStatistik/2017/02/Verdienstunterschiede_022017.pdf (last accessed 10 December 2018).

[11] Dumpert, F. (2018). Abschlussbericht zum Projekt „Prüfung und Bewertung von Optionen zur Schätzung der Staatsbürgerschaft in der Verdienststrukturerhebung (VSE)". Unpublished report.
Dumpert, F., Beck, M. (2017). Einsatz von Machine-Learning-Verfahren in amtlichen Unternehmensstatistiken. AStA Wirtschafts- und Sozialstatistisches Archiv, 11, 83–106, https://link.springer.com/content/pdf/10.1007%2Fs11943-017-0208-6.pdf (last accessed 10 December 2018).

| Institution | Project name | Description of the project | Application | Status | Method | Software |
|---|---|---|---|---|---|---|
| Federal Statistical Office of Germany | Transfer of the "minimum wage" property to the Integrated Employment Biographies (IEB) data[12] | The cross sectional Structure of Earnings Survey (SES) has information on the hourly wages and therefore also on whether employees are affected by the German minimum wage. This data is missing in the Integrated Employment Biographies (IEB), a panel provided by the Federal Employment Agency (BA/IAB). The idea is to enrich the panel data with the information on the minimum wage affection by training a random forest with the SES data and then applying it to the panel data. | Binary classification | Ongoing test | Random Forest | R |
| Federal Statistical Office of Germany | Machine Learning Methodology[13] | Evaluation of machine learning processes as a possibility for implementing automated data preparation and data analysis | Various | Tests | kNN, Naive Bayes, Random Forest, SVM, ANN, … | R, Python, HoloClean |
| Federal Statistical Office of Germany | classification of online job advertisements | Scraped online job advertisements are usually available in unstructured full text. There is often a lack of separate information, such as the economic sector, the desired level of education or whether the advertisement was placed by a recruiter or not. An ML approach should automatically carry out such classifications. | Text classification | Test | kNN, Multinomial-Naive-Bayes | Python (NLTK, SKlearn), R |
| Federal Statistical Office of Germany | Use of scanner data in consumer price statistics | Assignment of articles via their product descriptions to the ECOICOP classification. Eurostat has had a tool programmed which is to be adapted to the German situation. | Classification | Test | Three procedures are available in the tool, which are to be tested n: SVM, Random Forest, Logistic Regression. | The tool is programmed in Java and Ember.js as a web service. |

---

[12] Himmelreicher, R. K., vom Berge, P., Fitzenberger, B., Günther, R., Müller, D. (2017). Überlegungen zur Verknüpfung von Daten der Integrierten Erwerbsbiographien (IEB) und der Verdienststrukturerhebung (VSE). RatSWD Working Papers 262/2017, https://www.ratswd.de/dl/RatSWD_WP_262.pdf (last accessed 10 December 2018).
Dumpert, F., Beck, M. (2017). Einsatz von Machine-Learning-Verfahren in amtlichen Unternehmensstatistiken. AStA Wirtschafts- und Sozialstatistisches Archiv, 11, 83–106, https://link.springer.com/content/pdf/10.1007%2Fs11943-017-0208-6.pdf (last accessed 10 December 2018).
[13] Spies, L., Lange, K. (2018). Implementation of artificial intelligence and machine learning methods within the Federal Statistical Office of Germany. Working Paper for the UNECE Workshop on Statistical Data Editing September 2018, https://www.unece.org/fileadmin/DAM/stats/documents/ece/ces/ge.44/2018/T4_Germany_LANGE_Paper.pdf (last accessed 10 December 2018).

| Institution | Project name | Description of the project | Application | Status | Method | Software |
|---|---|---|---|---|---|---|
| Federal Statistical Office of Germany | Outlier Identification: Isolation Forest | Within the project "Implementation of the EU definition of the enterprise in business structure statistics", a donor-based imputation procedure is used for data collection. In order to remove insufficiently plausibilised or "extreme" data points from the donor pool, parametric outlier identification methods as well as an ML-based method were tested. Isolation Forest offers low effort in implementation and high efficiency in computing power even when dealing with large (structured) data sets. | Outlier identification | Test | Isolation Forest | R / Package: isofor;  Python / Package: Scikit-learn |
| Federal Statistical Office of Germany | Proof of Concept automated plausibility check (in the earnings statistics) | The use of the HoloClean software for automated plausibility checks of earnings statistics data is to be tested. HoloClean has an error and outlier detection module and learns a probability model based on several data sources that imputates (repairs) the erroneous data. | Error detection and imputation | Experiment | HoloClean (underlying probability model as factor graph, inference via Gibbs sampling) | Python (PyTorch), Apache Spark, PostgreSQL |
| Federal Statistical Office of Germany | Check the coding of part-time employment as part of the SV key of the BA (Federal Employment Agency) in the Structure of Earnings Survey (VSE). | The SV key of the BA, in which part-time employment is also coded, is also collected in the VSE. Erroneous information is corrected within the framework of plausibility checks on the basis of information on working time. Objective: With the corrected values, a model for predicting part-time work is learned and then applied to the erroneous BA data. The misclassifications then provide indications of cases in which part-time employment is incorrectly coded in the BA data. | Binary classification | Test | Random Forest | R (Ranger) |

| Institution | Project name | Description of the project | Application | Status | Method | Software |
|---|---|---|---|---|---|---|
| RKI / ZBS 6 | Identification, differentiation and classification of biological samples such as cells, tissues and microorganisms based on spectroscopic and spectrometric data | The ZBS 6 department of the RKI has been using methods of machine learning and artificial intelligence (AI) for more than 20 years. Among other things, variants of artificial neural networks (ANNs) are used, which enable pattern recognition based on spectral data, e.g. from vibration spectroscopy or mass spectrometry. Spectroscopic and spectrometric data are recorded from complex biological samples such as cells, tissues, etc.; the aim of the application of ANNs is the fast, objective and cost-effective identification, differentiation and classification in cytology, histology and microbiology within the framework of research activities or for method development in diagnostics. | Differentiation, identification, classification | Research projects | ANN SVM Strategies for Optimization | NeuroDeveloper Matlab Matlab NN Toolbox Biotools ga_orstooldiag SNNS |
| RKI / MF1 | Bioinformatics / Analysis of molecular data[14] | Various methods of machine learning are developed and used for the analysis of large data sets from omics experiments such as genome sequencing, automated assistance systems. These data sets are so large (sometimes up to one billion genome fragments from a single experiment) that manual analysis is not fully useful or possible, especially for time-critical processes. This can be used, for example, to characterise bacterial or viral pathogens or to search for pathogens. The assistance systems are used in particular to identify irrelevant measurements for the human decision-maker, e.g. genome fragments that were only measured with low quality, that have biologically low information content or contain contaminations from measurement, environment or a host organism. In addition, special hazard potentials of individual measurements (e.g. proximity to known pathogens or relevant potential phenotypes such as virulence or resistance) for the human decision-maker are highlighted. | Classification (binary and multiclass problems), regression, clustering | Research projects and productive operation | Random Forests, Deep Learning | R, Python |

---

[14] Deneke, C., Rentzsch, R., Renard, B. Y. (2017). PaPrBaG: A machine learning approach for the detection of novel pathogens from NGS data. Scientific Reports, 7, No. 39194, https://www.nature.com/articles/srep39194 (last accessed 10 December 2018).
Löwer, M., Renard, B. Y., de Graaf, J., Wagner, M., Paret, C., Kneip, C., Türeci, Ö., Diken, M., Britten, C., Kreiter, S., Koslowski, M., Castle, J. C., Sahin, U. (2012). Computational Biology, September 2012, https://doi.org/10.1371/journal.pcbi.1002714 (last accessed 10 December 2018).
Hanselmann, M., Röder, J., Köthe, U., Renard, B. Y., Heeren, R. M. A., Hamprecht, F. A. (2013). Analytical Chemistry, 85(1), 147–155, https://pubs.acs.org/doi/abs/10.1021/ac3023313 (last accessed 10 December 2018).
Kuhring, M., Dabrowski, P. W., Piro, V. C., Nitsche, A., Renard, B. Y. (2015). BMC Bioinformatics, 16(1), 240, https://doi.org/10.1186/s12859-015-0644-7 (last accessed 10 December 2018).

Martin Beck; Florian Dumpert; Joerg Feuerhake: Machine Learning in Official Statistics| Institution | Project name | Description of the project | Application | Status | Method | Software |
|---|---|---|---|---|---|---|
| RKI / FG-31 | Automatic outbreak detection for infectious diseases[15] | We develop and implement methods of machine learning for the prediction of case numbers of infectious diseases and for the detection of anomalies in surveillance data. We also use supervised learning approaches to compare and optimize these algorithms. Within the "Signals 2.0" project further application research concerning Natural Language Processing for the processing of unstructured data (e.g. protocols, publications) is planned, the extracted information should be made available and also included in the outbreak detection. | Regression, classification and information retrieval | Research projects and productive operation | HMM, GLM, NLP, classification algorithms (Logistic Regression, Random Forest, SVM, Neural Networks) | R, Python |
| RKI / P4 | Application of machine learning methods in RKI health monitoring (KiGGS)[16] | As part of the KiGGS study, the RKI collected around 4000 variables on the health of over 17,000 children and young adults. Due to the size and heterogeneity of the data set, classical statistical methods sometimes reach conceptual limits. As a pilot project for the use of machine learning methods in public health monitoring, this project is intended to show how novel methods of machine learning can be used in the work of the RKI. Various methods will be tested to identify structures and correlations in the data and to predict the occurrence of chronic diseases. | Clustering, binary and multi-class classification | Experiment | Cluster methods, Decision Trees, Neural Networks | Python |

---

[15] https://www.rki.de/signale-projekt (last accessed 10 December 2018).
Salmon, M., Schumacher, D., Burmann, H., Frank, C., Claus, H., Höhle, M. (2016). A system for automated outbreak detection of communicable diseases in Germany. Eurosurveillance, 21(13), 30180, http://dx.doi.org/10.2807/1560-7917.ES.2016.21.13.30180 (last accessed 10 December 2018).
[16] https://www.kiggs-studie.de/ (last accessed 10 December 2018).

| Institution | Project name | Description of the project | Application | Status | Method | Software |
|---|---|---|---|---|---|---|
| Centre for European Economic Research (ZEW) | TOBI - Text Data-Based Output Indicators as the basis for a new innovation metric[17] | New output indicators for innovation activities are being developed in the joint project. Computational linguistic methods are used, which are applied to large amounts of text data. The development of the methods and validation of the generated indicators is carried out by the ZEW in Mannheim and the Justus-Liebig-University Gießen. At the ZEW, the analysis is based on text content from company websites, which is collected automatically and regularly via a web scraper. Using text data mining (e.g. topic models), information on innovations is then identified from these texts and innovation indicators are derived. The websites are accessed on the basis of the databases available at the ZEW. These allow the continuous monitoring of the websites of the current German company stock and the consideration of extensive metadata (e.g. industry and location of the company). In addition, the newly generated innovation indicators can be compared with conventional innovation indicators via the ZEW databases. | Text analysis and classification | Under development | Miscellaneous | Python |
| Centre for European Economic Research (ZEW) | Science4KMU[18] | Assessment of the willingness of companies to cooperate using a neural network based on survey data from the Community Innovation Survey (CIS). The model is designed to make the contact between Technology Transfer Offices (TTO) and companies more efficient. | Probabilistic scoring model | Prototype | Neural Network | Stata |
| Hessian Statistical Office | Webscraping of enterprise websites[19] | Web scraping of enterprise websites and machine learning to gain new digital data | Data extraction, coherence checks | Test mode | Webscraping, graph-based neural networks | Java, R, MySql, Apache Spark |

---

[17] http://www.zew.de/de/forschung/textdaten-basierte-output-indikatoren-als-basis-einer-neuen-innovationsmetrik/ (last accessed 10 December 2018).
[18] http://www.zew.de/de/forschung/innovationswerkstatt-leibniz-entwicklung-der-innovationsfaehigkeit-von-forschungsinstitutionen-im-querschnittsbereich-durch-einbindung-von-kmu-science4kmu/?cHash=58bc9e70f383efcf1bc5409edfe7521fscience4kmu/?cHash=58bc9e70f383efcf1bc5409edfe7521f (last accessed 10 December 2018).
[19] Hessisches Statistisches Landesamt (2018). Webscraping von Unternehmenswebseiten und maschinelles Lernen zum Gewinnen von neuen digitalen Daten. Report, https://statistik.hessen.de/sites/statistik.hessen.de/files/Webscraping_von_Unternehmenswebseiten.pdf (last accessed 10 December 2018).

| Institution | Project name | Description of the project | Application | Status | Method | Software |
|---|---|---|---|---|---|---|
| German Central Bank | Statistics on securities investments: Support of data preparation through integration of machine learning procedures[20] | A Random Forest algorithm learns to mark data reported from employee decisions as incorrect and to make predictions about the probability of error. | Classification | Prototype | Random Forest | Python |
| German Central Bank | Record Linkage in FDSZ[21] | Identification of units from various internal and external company databases for the creation of linked company data for research purposes. | Classification | Prototypes | Random Forest | Python, SAS |
| German Central Bank | Combination of statistical tests for the existence of seasonal patterns in macroeconomic time series[22] | The random forest algorithm, which can be assigned to machine learning, is repeatedly one of the best forecasting and classification methods. At the same time, measures exist for determining the importance of predictors. Thus, random forests can be used to combine the results of different seasonal tests and quantify the influence of each test on the final classification decision and thus the information content of each test. | Classification | Project | Random Forest, Conditional Random Forest | R |

---

| Institution | Project name | Description of the project | Application | Status | Method | Software |
|---|---|---|---|---|---|---|
| German Central Bank | Classification of economic activities for sole proprietorships[23] | Economic activities are provided by various data providers in the non-financial individual accounts statistics of the Deutsche Bundesbank. The quality of these sources varies over time as well as in cross-section. A poor allocation of the branch of economic activity can lead to distortions in the company accounts statistics, in particular by holding companies.<br><br>As holding companies differ from other companies in their financial structure, it is classified whether a company is a holding company or not. Doubtful cases are then examined manually. | Classification | Implemented | Artificial neural backpropagation network with an error learning sequence from Support Vector Machine and Random Forest. (previously Logit Backward Selection) | R (formerly Stata) |
| IAB FB B2 | Typification of labour market regions (comparison types SGB III / SGB II)[24] | In the second stage of the typification process, employment agency or job centre districts with similar regional labour market characteristics are clustered through the use of Unsupervised Learning. The Ward method is combined with a k-Means algorithm. | Clustering | Productive operation | Clustering according to Ward / k-Means algorithm | Stata |
| IAB & University of Mannheim | New methods for professional coding[25] | Professional code refers to the assignment of free text answers from surveys to official professional classifications. The generation of proposals for manual coding has so far been partly automatic, based on a list of job designations and string matching algorithms. The project is testing whether training data from previous surveys can be used as an alternative to generate better proposals. Furthermore, the suggestions should be displayed directly during the interview so that the interviewees can choose the most suitable category themselves. | Classification with several classes | Under development | Comparison of different algorithms, Stacking, Boosting | R |
| IAB | Correction of training information in administrative data | Within the framework of the project, a data-driven procedure for the correction of educational information in administrative individual data is to be developed and compared with current deterministic methods. | Classification with several classes | Under development | Decision Trees, Random Forest, SVM, Boosting, Ensembles, Stacking | R |

---

[23] http://www.statistische-woche.de/fileadmin/Rostock/BookofAbstracts_Rostock.pdf (pp.165-166) (last accessed 10 December 2018).
[24] Blien, U., Hirschenauer, F., Phan thi Hong, V. (2010). Classification of regional labour markets for purposes of labour market policy. Papers in Regional Science, 89(4), 859–880, https://onlinelibrary.wiley.com/doi/abs/10.1111/j.1435-5957.2010.00331.x (last accessed 10 December 2018).
[25] Schierholz, M., Gensicke, M., Tschersich, N., Kreuter, F. (2017). Occupation coding during the interview. Journal of the Royal Statistical Society: Series A, 181(2), 379–407, http://dx.doi.org/10.1111/rssa.12297 (last accessed 10 December 2018).

| Institution | Project name | Description of the project | Application | Status | Method | Software |
|---|---|---|---|---|---|---|
| IAB | Prediction of duration in unemploy-ment | In the course of the project, the duration of unemployment is to be predicted by customers of the Federal Employment Agency. The project is explorative. The preliminary goal is therefore to evaluate the potential of machine learning in this context. | Regression | Idea / Feasibility study | Decision Trees, Random Forest, SVM, Neural Networks, Boosting, Ensembles, Stacking | R |
| IAB | Prediction of duration in unemployment and evaluation of placement measures on the basis of a field experiment with machine learning methods[26] | Within the framework of the DFG project, context-related possibilities for predicting unemployment with machine learning will be tested and used to evaluate the effects of a field experiment. | Regression and classification | Ongoing DFG project | Decision Trees, Random Forest, SVM, Boosting, Ensembles, Stacking | R |
| IAB | Estimation of economic activities using CART procedures | Regular changes in the classification of economic activities cause problems when evaluating over longer periods of time. As part of the project, individual classifications will be updated for the IAB's operational history panel over the entire period of the panel, so that evaluations will be possible on a consistent classification. | Classification with several classes | Under development | Decision Trees | R |
| IAB FB C1 and Bristol University | Prediction of unemploy-ment duration[27] | The aim of the project is to predict unemployment by means of machine learning methods and to investigate already conducted experiments on different aspects of assignment processes to heterogeneous treatment effects, also with reference to machine learning methods. | Classification and prediction | Productive operation | Decision Trees, Logit, Lasso, SVM | R |

---

[26] http://gepris.dfg.de/gepris/projekt/387482412 (last accessed 10 December 2018).
[27] http://gepris.dfg.de/gepris/projekt/387482412 (last accessed 10 December 2018).

| Institution | Project name | Description of the project | Application | Status | Method | Software |
|---|---|---|---|---|---|---|
| IAB | Imputation of missing information on working time | As a result of the change in the activity characteristic of the social security declarations (DEÜV), the full-time/part-time specification was missing in the employee declarations in 2011 and 2012. They were imputed with the help of classification trees. | Classification | Completed | Classification Trees, Cross Validation | R |
| Federal Office for Migration and Refugees | Profile analysis | With the implementation of a proof of concept and the subsequent development of a pilot system, the profile analysis project will test machine learning methods at the Federal Office for Migration and Refugees for the first time. A semantic text analysis of the hearing records in the context of the asylum procedure is provided. The aim is to highlight relevant text passages that point to safety-relevant information and to classify them according to given safety criteria. The analysis results are prepared in a user-friendly way and thus support the responsible clerks in fulfilling the BAMF's reporting obligation. | Aggregation and classification of text passages | Development of a pilot system | Semantic text analysis | Watson Explorer (WEX) |
| GESIS - Leibniz Institute for the Social Sciences | Election Campaigning on Social Media: Politicians, Audiences and the Mediation of Political Communication on Facebook and Twitter[28] | Abstract: "Although considerable research has concentrated on online campaigning, it is still unclear how politicians use different social media platforms in political communication. Focusing on the German federal election campaign 2013, this article investigates whether election candidates address the topics most important to the mass audience and to which extent their communication is shaped by the characteristics of Facebook and Twitter. Based on open-ended responses from a representative survey conducted during the election campaign, we train a human-interpretable Bayesian language model to identify political topics. Applying the model to social media messages of candidates and their direct audiences, we find that both prioritize different topics than the mass audience. The analysis also shows that politicians use Facebook and Twitter for different purposes. We relate the various findings to the mediation of political communication on social media induced by the particular characteristics of audiences and sociotechnical environments." | Topic Modelling | Research publication | Semi-supervised classification | Juliet |

---

[28] Stier, S., Bleier, A., Lietz, H., Strohmaier, M. (2018). Election Campaigning on Social Media: Politicians, Audiences, and the Mediation of Political Communication on Facebook and Twitter. Political Communication, 35(1), 50–74, https://doi.org/10.1080/10584609.2017.1334728 (last accessed 10 December 2018).

| Institution | Project name | Description of the project | Application | Status | Method | Software |
|---|---|---|---|---|---|---|
| GESIS - Leibniz Institute for the Social Sciences | When populists become popular: comparing Facebook use by the right-wing movement Pegida and German political parties[29] | Abstract: "Previous research has acknowledged the use of social media in political communication by right-wing populist parties and politicians. Less is known, however, about its pivotal role for right-wing social movements which rely on personalized messages to mobilize supporters and challenge the mainstream party system. This paper analyses online political communication by the right-wing populist movement Pegida and German political parties. We investigate to which extent parties attract supporters of Pegida, to which extent they address topics similar to Pegida and whether their topic use has become more similar over a period of almost two years. The empirical analysis is based on Facebook posts by main accounts and individual representatives of these political groups. We first show that there are considerable overlaps in the audiences of Pegida and the new challenger in the party system, AfD. Then we use topic models to characterize topic use by party and surveyed crowd workers to which extent they perceive the identified topics as populist communication. The results show that while Pegida and AfD talk about rather unique topics and smaller parties engage to varying degrees with the topics populists emphasize, the two governing parties CDU and SPD clearly deemphasize those. Overall, the findings indicate that the considerable attention devoted to populist actors and shifts in public opinion due to the refugee crisis have left only moderate marks in political communication within the mainstream party system." | Topic Modelling | Research publication | LDA | Python |

---

[29] Stier, S., Posch, L., Bleier, A., Strohmaier, M. (2017). When populists become popular: comparing Facebook use by the right-wing movement Pegida and German political parties. Information, Communication & Society, 20(9), 1365–1388, https://doi.org/10.1080/1369118X.2017.1328519 (last accessed 10 December 2018).

| Institution | Project name | Description of the project | Application | Status | Method | Software |
|---|---|---|---|---|---|---|
| GESIS - Leibniz Institute for the Social Sciences | Towards Quantifying Sampling Bias in Network Inference[30] | Abstract: "Relational inference leverages relationships between entities and links in a network to infer information about the network from a small sample. This method is often used when global information about the network is not available or difficult to obtain. However, how reliable is inference from a small labelled sample? How should the network be sampled, and what effect does it have on inference error? How does the structure of the network impact the sampling strategy? We address these questions by systematically examining how network sampling strategy and sample size affect accuracy of relational inference in networks. To this end, we generate a family of synthetic networks where nodes have a binary attribute and a tunable level of homophily. As expected, we find that in heterophilic networks, we can obtain good accuracy when only small samples of the network are initially labelled, regardless of the sampling strategy. Surprisingly, this is not the case for homophilic networks, and sampling strategies that work well in heterophilic networks lead to large inference errors. These findings suggest that the impact of network structure on relational classification is more complex than previously thought." | Relational Classification | Research publication | Bayes + Relaxation + Collective Inference | Python |

---

[30] Espín-Noboa, L., Wagner, C., Karimi, F., Lerman, K. (2018). Towards Quantifying Sampling Bias in Network Inference. Cornell University arXiv:1803.02422, https://arxiv.org/abs/1803.02422 (last accessed 10 December 2018).

| Institution | Project name | Description of the project | Application | Status | Method | Software |
|---|---|---|---|---|---|---|
| GESIS - Leibniz Institute for the Social Sciences | How Users Explore Ontologies on the Web: A Study of NCBO's BioPortal Usage Logs[31] | Abstract: "Ontologies in the biomedical domain are numerous, highly specialized and very expensive to develop. Thus, a crucial prerequisite for ontology adoption and reuse is effective support for exploring and finding existing ontologies. Towards that goal, the National Center for Biomedical Ontology (NCBO) has developed BioPortal---an online repository containing more than 500 biomedical ontologies. In 2016, BioPortal represents one of the largest portals for exploration of semantic biomedical vocabularies and terminologies, which is used by many researchers and practitioners. While usage of this portal is high, we know very little about how exactly users search and explore ontologies and what kind of usage patterns or user groups exist in the first place. Deeper insights into user behavior on such portals can provide valuable information to devise strategies for a better support of users in exploring and finding existing ontologies, and thereby enable better ontology reuse. To that end, we study and group users according to their browsing behavior on BioPortal and use data mining techniques to characterize and compare exploration strategies across ontologies. In particular, we were able to identify seven distinct browsing types, all relying on different functionality provided by BioPortal. For example, Search Explorers extensively use the search functionality while Ontology Tree Explorers mainly rely on the class hierarchy for exploring ontologies. Further, we show that specific characteristics of ontologies influence the way users explore and interact with the website. Our results may guide the development of more user-oriented systems for ontology exploration on the Web." | Clustering, dimensionality reduction | Research publication | k-Means + PCA | Python, R |
| GESIS - Leibniz Institute for the Social Sciences | Practical collapsed stochastic variational inference for the HDP[32] | In this work we explore a collapsed stochastic variational Bayes inference for the Hierarchical Dirichlet process (HDP). The proposed online algorithm is easy to implement and accounts for the inference of hyper-parameters. | Bayesian non-parametric mixed-membership clustering | Research publication | PCSVB0 | Juliet |

---

[31] Walk, S., Esín-Noboa, L., Helic, D., Strohmaier, M., Musen, M. A. (2017). How Users Explore Ontologies on the Web: A Study of NCBO's BioPortal Usage Logs. Proceedings of the 26th International Conference on World Wide Web, 775–784, https://doi.org/10.1145/3038912.3052606 (last accessed 10 December 2018).
[32] Bleier, A. (2013). Practical Collapsed Stochastic Variational Inference for the HDP. Cornell University arXiv:1312.0412, http://arxiv.org/abs/1312.0412 (last accessed 10 December 2018).

| Institution | Project name | Description of the project | Application | Status | Method | Software |
|---|---|---|---|---|---|---|
| GESIS - Leibniz Institute for the Social Sciences | Polylingual Labeled Topic Model[33] | Development and evaluation of the Polylingual Labeled Topic Model | Topic Model | Research publication | PLL-TM | Julia, CML |
| GESIS - Leibniz Institute for the Social Sciences | Predicting structured metadata from unstructured metadata[34] | Framework to predict structured metadata terms from unstructured metadata for improving quality and quantity of metadata, using the Gene Expression Omnibus (GEO) microarray database | Metadata prediction | Research publication | LDA, SVM | Scala, Python |
| GESIS - Leibniz Institute for the Social Sciences | Enriching ontologies with encyclopedic background knowledge for document indexing[35] | Using encyclopedic background knowledge for enriching domain-specific ontologies for document classification | Document classification | Research publication | LLDA, SVM | Scala, Python |
| GESIS - Leibniz Institute for the Social Sciences | Measuring Motivations ofCrowd-workers: The Multidimensional Crowd-worker Motivation Scale[36] | Presents the Multidimensional Crowdworker Motivation Scale (MCMS), a scale for measuring the motivation of crowdworkers on micro-task platforms. | Scale development | Research | Structural Equation Models | R, MPlus, Python, CML |

---

[33] Posch, L., Bleier, A., Schaer, P., Strohmaier, M. (2015). The Polylingual Labeled Topic Model. Cornell University arXiv:1507.06829, https://arxiv.org/abs/1507.06829 (last accessed 10 December 2018).
[34] Posch, L., Panahiazar, M., Dumontier, M., Gevaert, O. (2016). Predicting structured metadata from unstructured metadata. Database, 2016, baw080, https://academic.oup.com/database/article/doi/10.1093/database/baw080/2630448 (last accessed 10 December 2018).
[35] Posch, L. (2016). Enriching Ontologies with Encyclopedic Background Knowledge for Document Indexing. Cornell University arXiv: 1603.06494, https://arxiv.org/pdf/1603.06494.pdf (last accessed 10 December 2018).
[36] Posch, L., Bleier, A., Strohmaier, M. (2017). Measuring Motivations of Crowdworkers: The Multidimensional Crowdworker Motivation Scale. Cornell University arXiv: 1702.01661, https://arxiv.org/pdf/1702.01661.pdf (last accessed 10 December 2018).

| Institution | Project name | Description of the project | Application | Status | Method | Software |
|---|---|---|---|---|---|---|
| GESIS - Leibniz Institute for the Social Sciences | A System for Probabilistic Linking of Thesauri and Classification Systems[37] | Presents a system which creates and visualizes probabilistic semantic links between concepts in a thesaurus and classes in a classification system. | Concept linking | Research | PLL-TM | Julia, CML, D3 |
| GESIS - Leibniz Institute for the Social Sciences | Why we read Wikipedia[38] | Abstract: "Wikipedia is one of the most popular sites on the Web, with millions of users relying on it to satisfy a broad range of information needs every day. Although it is crucial to understand what exactly these needs are in order to be able to meet them, little is currently known about why users visit Wikipedia. The goal of this paper is to fill this gap by combining a survey of Wikipedia readers with a log-based analysis of user activity. Based on an initial series of user surveys, we build a taxonomy of Wikipedia use cases along several dimensions, capturing users' motivations to visit Wikipedia, the depth of knowledge they are seeking, and their knowledge of the topic of interest prior to visiting Wikipedia. Then, we quantify the prevalence of these use cases via a large-scale user survey conducted on live Wikipedia with almost 30,000 responses. Our analyses highlight the variety of factors driving users to Wikipedia, […]. Finally, we match survey responses to the respondents' digital traces in Wikipedia's server logs, enabling the discovery of behavioral patterns associated with specific use cases. For instance, we observe long and fast-paced page sequences across topics for users who are bored or exploring randomly, whereas those using Wikipedia for work or school spend more time on individual articles focused on topics such as science. Our findings advance our understanding of reader motivations and behavior on Wikipedia and can have implications for developers aiming to improve Wikipedia's user experience, editors striving to cater to their readers' needs, third-party services (such as search engines) providing access to Wikipedia content, and researchers aiming to build tools such as recommendation engines." | Classification | Research publication | Gradient Boosting | Python, Spark |

---

[37] Posch, L., Schaer, P., Bleier, A., Strohmaier, M. (2016). A System for Probabilistic Linking of Thesauri and Classification Systems. KI – Künstliche Intelligenz, 30(2), 193–196, https://link.springer.com/article/10.1007/s13218-015-0413-9 (last accessed 10 December 2018).
[38] Singer, P., Lemmerich, F., West, R., Zia, L., Wulczyn, E., Strohmaier, M., Leskovec, J. (2017). Why We Read Wikipedia. Cornell University arXiv: 1702.05379, https://arxiv.org/abs/1702.05379 (last accessed 10 December 2018).

| Institution | Project name | Description of the project | Application | Status | Method | Software |
|---|---|---|---|---|---|---|
| GESIS - Leibniz Institute for the Social Sciences | Predicting Genre Preferences from Cultural and Socio-economic Factors for Music Retrieval[39] | Abstract: "In absence of individual user information, knowledge aboutlarger user groups (e.g., country characteristics) can be exploited forderiving user preferences in order to provide recommendations to users.In this short paper, we study how to mitigate the cold-start problem on acountry level for music retrieval. Specifically, we investigate a large-scaledataset on user listening behavior and show that we can reduce the errorfor predicting the popularity of genres in a country by about 16.4% overa baseline model using cultural and socio-economics indicators." | Regression | Research publication | Gradient Boosting, Random Forest | Python |
| GESIS - Leibniz Institute for the Social Sciences | Discovering and Characterizing Mobility Patterns in Urban Spaces: A Study of Manhattan Taxi Data[40] | Abstract: "Nowadays, human movement in urban spaces can be traced digitallyin many cases. It can be observed that movement patternsare not constant, but vary across time and space. In this work,we characterize such spatio-temporal patterns with an innovativecombination of two separate approaches that have been utilized forstudying human mobility in the past. First, by using non-negativetensor factorization (NTF), we are able to cluster human behaviorbased on spatio-temporal dimensions. Second, for characterizingthese clusters, we propose to use HypTrails, a Bayesian approachfor expressing and comparing hypotheses about human trails. Toformalize hypotheses, we utilize publicly available Web data (i.e.,Foursquare and census data). By studying taxi data in Manhattan,we can discover and characterize human mobility patterns that cannotbe identified in a collective analysis. As one example, we finda group of taxi rides that end at locations with a high number ofparty venues on weekend nights. Our findings argue for a morefine-grained analysis of human mobility in order to make informeddecisions for e.g., enhancing urban structures, tailored traffic controland location-based recommender systems." | Clustering | Research publication | Tensor Factorization | Python |

---

[39] Skowron, M., Lemmerich, F., Ferwerda, B., Schedl, M. (2017). ECIR 2017: Advances in Information Retrieval, 561–567, https://link.springer.com/chapter/10.1007/978-3-319-56608-5_49 (last accessed 10 December 2018).
[40] Espín-Noboa, L., Lemmerich, F., Singer, P., Strohmaier, M. (2016). Discovering and Characterizing Mobility Patterns in Urban Spaces: A Study of Manhattan Taxi Data. Proceedings of the 25th International Conference Companion on World Wide Web, 537–542, http://gdac.uqam.ca/WWW2016-Proceedings/companion/p537.pdf (last accessed 10 December 2018).

| Institution | Project name | Description of the project | Application | Status | Method | Software |
|---|---|---|---|---|---|---|
| GESIS - Leibniz Institute for the Social Sciences | Extracting Semantics from Random Walks on Wikipedia: Comparing Learning and Counting Methods.[41] | Abstract: "Semantic relatedness between words has been extracted from a variety of sources. In this ongoing work, we explore and compare several options for determining if semantic relatedness can be extracted from navigation structures in Wikipedia. In that direction, we first investigate the potential of representation learning techniques such as DeepWalk in comparison to previously applied methods based on counting co-occurrences. Since both methods are based on (random) paths in the network, we also study different approaches to generate paths from Wikipedia link structure. For this task, we do not only consider the link structure of Wikipedia, but also actual navigation behavior of users. Finally, we analyze if semantics can also be extracted from smaller subsets of the Wikipedia link network. As a result we find that representa- tion learning techniques mostly outperform the investigated co-occurrence counting methods on the Wikipedia network. However, we find that this is not the case for paths sampled from human navigation behavior." | Semantic relatedness | Research publication | Deep Learning | Python |

---

[41] Dallmann, A., Niebler, T., Lemmerich, F., Hotho, A. (2016, April). Extracting Semantics from Random Walks on Wikipedia: Comparing Learning and Counting Methods. In Wiki@ ICWSM, https://www.aaai.org/ocs/index.php/ICWSM/ICWSM16/paper/download/13199/12869 (last accessed 10 December 2018).

| Institution | Project name | Description of the project | Application | Status | Method | Software |
|---|---|---|---|---|---|---|
| GESIS - Leibniz Institute for the Social Sciences | Text Categorization for Deriving the Application Quality in Enterprises Using Ticketing Systems[42] | Abstract: "Today's enterprise services and business applications are often centralized in a small number of data centers. Employees located at branches and side offices access the computing infrastructure via the internet using thin client architectures. The task to provide a good application quality to the employers using a multitude of different applications and access networks has thus become complex. Enterprises have to be able to identify resource bottlenecks and applications with a poor performance quickly to take appropriate countermeasures and enable a good application quality for their employees. Ticketing systems within an enterprise use large databases for collecting complaints and problems of the users over a long period of time and thus are an interesting starting point to identify performance problems. However, manual categorization of tickets comes with a high workload. In this paper, we analyze in a case study the applicability of supervised learning algorithms for the automatic identification of relevant tickets, i.e., tickets indicating problematic applications. In that regard, we evaluate different classification algorithms using 12,000 manually annotated tickets accumulated in July 2013 at the ticketing system of a nation-wide operating enterprise. In addition to traditional machine learning metrics, we also analyze the performance of the different classifiers on business-relevant metrics." | Text classification | Research publication | SVM, Decision Tree and others | Java |

---

[42] Zinner, T., Lemmerich, F., Schwarzmann, S., Hirth, M., Karg, P., Hotho, A. (2015). Text Categorization for Deriving the Application Quality in Enterprises Using Ticketing Systems. DaWaK 2015: Big Data Analytics and Knowledge Discovery, 325–336, https://link.springer.com/chapter/10.1007/978-3-319-22729-0_25 (last accessed 10 December 2018).

| Institution | Project name | Description of the project | Application | Status | Method | Software |
|---|---|---|---|---|---|---|
| GESIS - Leibniz Institute for the Social Sciences | RDF Vocabulary Term Recommen-dation[43] | Abstract: "Deciding which RDF vocabulary terms to use when modeling data as Linked Open Data (LOD) is far from trivial. In this paper, we propose TermPicker as a novel approach enabling vocabulary reuse by recommending vocabulary terms based on various features of a term. These features include the term's popularity, whether it is from an already used vocabulary, and the so-called schema-level pattern (SLP) feature that exploits which terms other data providers on the LOD cloud use to describe their data. We apply Learning To Rank to establish a ranking model for vocabulary terms based on the utilized features. The results show that using the SLP-feature improves the recommendation quality by 29–36 % considering the Mean Average Precision and the Mean Reciprocal Rank at the first five positions compared to recommendations based on solely the term's popularity and whether it is from an already used vocabulary." | Learning To Rank | Research publication | Various Learning To Rank algorithms from the RankLib library | RankLib Library (Java) |

---

[43] Schaible, J., Gottron, T., Scherp, A. (2016). TermPicker: Enabling the Reuse of Vocabulary Terms by Exploiting Data from the Linked Open Data Cloud. ESWC 2016: The Semantic Web. Latest Advances and New Domains, 101–117, https://link.springer.com/chapter/10.1007/978-3-319-34129-3_7 (last accessed 10 December 2018).
Schaible, J., Gottron, T., Scherp, A. (2015). TermPicker: enabling the reuse of vocabulary terms by exploiting data from the linked open data cloud – an extended technical report. Cornell University arXiv: 1512.05685,https://arxiv.org/abs/1512.05685 (last accessed 10 December 2018).

## 4.2 Application of machine learning in international statistical institutions

| Institution | Project name | Description of the project | Application | Status | Method | Software |
|---|---|---|---|---|---|---|
| Australian Bureau of Statistics | Automated coding for complex classifications | | Multilevel multiclass classification | Productive | SVM, Bootstrap Aggregation, text models | C, Java |
| Australian Bureau of Statistics | Linking and grouping of statistical entities by their pattern of connections | | Graph structure classification, motif detection | Experiment | SVM, Graph Kernel Models | R |
| Australian Bureau of Statistics | Land utilisation and crop prediction | | Multiclass classification | Experiment | CNN | R |
| Australian Bureau of Statistics | Editing administrative datasets | | Multiclass prediction, anomaly detection | Experiment | Probabilistic Graphical Models | R |
| Australian Bureau of Statistics | Predicting Census occupancy | | Binary classification | Experiment | Decision Tree | R |
| Australian Bureau of Statistics | Linking multiple datasets | | Entity resolution creation of a linking spine | Experiment | Empirical Bayes | R, Scala |

| Institution | Project name | Description of the project | Application | Status | Method | Software |
|---|---|---|---|---|---|---|
| Central Statistical Bureau of Latvia | Number of population and key demographic indicators[44] | | Binary classification | Productive | Logistic Regression (GLM) is used in production. Stochastic Gradient Boosting (GBM), Support Vector Machines (SVM), Regularized Discriminant Analysis (RDA), Multi-Layer Perceptron (MLP), Radial Basis Function Network (RBF) were tested during the implementation stage. | R (data.table) |
| Central Statistics Office of Ireland | Automatic coding via open-source indexing utility | An automatic coding system for Classification of Individual Consumption by Purpose (COICOP) assignment for their Household Budget Survey, using previously coded records as training data. Their method is based on the open-source indexing and searching tool Apache Lucene (http://lucene.apache.org). | Multi-class classification | Development | | Apache Lucene / Python |
| Eurostat | Euro Area GDP Forecast Using Large Survey Dataset - A Random Forest Approach | This paper presents a new statistical approach to forecasting macro-economic aggregates, based on the Random Forests technique, originally developed as a learning classification tool (Breiman, 2001). This technique can handle a very large number of input variables without overfitting and is known to enjoy good prediction properties and to be robust to noise. | Regression | Experiment | Random Forest | R |

---

[44] http://old.csb.gov.lv/en/statistikas-temas/metodologija/number-population-and-key-demographic-indicators-36808.html (last accessed 10 December 2018).

| Institution | Project name | Description of the project | Application | Status | Method | Software |
|---|---|---|---|---|---|---|
| Eurostat | Categorical data imputation via neural networks and Bayesian networks[45] | Eurostat compared imputation results for missing categorical data (voting intensions) based on two machine learning methods (neural networks and Bayesian networks) against one of the current prevailing statistical imputation methods (multiple imputation using logistic regression). | Imputation | Experiment | Logistic Regression, Neural Networks, Bayesian Networks | (SAS) |
| Federal Statistical Office of Switzerland | Modelling of the non-response mechanism | Classification trees are used to model the behaviour of the non-respondents in order to diminish non-response bias in the results. | Binary classification of response homogeneity groups | Productive | CHAID | SAS |
| Federal Statistical Office of Switzerland | Detection of suspicious responses | Several machine learning algorithms are tested to detect anomalies in the data. Detected units are contacted to check the data again. These methods are applied in a field where no or very few edit rules are available. | Binary classification | Test | Generalized boosted models, Random Forest, Neural networks, Naive Bayes, Tree algorithms, … | R |
| Federal Statistical Office of Switzerland | Turnover breakdown from grouped answers to the enterprises | This project is at its beginning. It is planned to use random forests to learn from the Turnover statistics and to apply it to the units not in the Turnover statistics. There is no other data available to breakdown grouped VAT data. | Classification | Test | Random Forest | R |
| Federal Statistical Office of Switzerland | Automatization of the NOGA (Swiss NACE) coding | This project is at its very beginning. It is planned to test several machine learning algorithms which are not yet fixed. | Classification | Test | Several | R |
| Federal Statistical Office of Switzerland | Automatization of land cover and land use codes based on aerial images | This project is at its beginning. It is planned to use convolutional neural networks for coding or for change detection. | Classification | Test | CNN | R/Python |

---

[45] https:/www.unece.org/fileadmin/DAM/stats/documents/ece/ces/ge.44/2012/Presentations/Topic_6__37__Eurostat.pptx (last accessed 10 December 2018).

| Institution | Project name | Description of the project | Application | Status | Method | Software |
|---|---|---|---|---|---|---|
| Federal Statistical Office of Switzerland | Clustering of the "careers" in the social security system | This project is at its beginning. It is planned to test whether it is possible to detect similar "careers" in the social security system automatically. | Clustering | Test | Several | R |
| Hungarian Central Statistical Office | Tax evader detection[46] | Detecting self-employed proprietors who are tax evaders | Classification | Experiment | k-NN | |
| Institut national de la statistique et des études économiques (INSEE) | Detecting wages/paid hours anomalies in employer payroll declaration statistical databases | The Annual Declaration of Social Data ("déclaration annuelle de données sociales", DADS), mandatory fulfilled each year by each employer and to which reported individual wage-earner information is transmitted to fiscal and social services for payroll and tax purposes as well as for calculating social security wage-earners rights (e.g., pensions), has been replaced since 2016 by a monthly Nominative Social Declaration information. This change of sources completely modifies the national statistical service of information on employment and wages that relies on, and provides the opportunity to rethink the automatic anomaly detection process implemented in the statistical production line. The experimental project tests different machine learning-based algorithms for anomaly detection of net and gross wages and related paid hours. | Non supervised algorithms | Ongoing experiment | Fuzzy associations, Isolation Forest, local outliers factors | R, Python |
| Institut national de la statistique et des études économiques (INSEE) | Nowcasting short-term business indicators with media information | Several experiments aimed to assess the performance of nowcasting different business indicators (e.g. GDP, payroll employment) with google trends, and traditional media information. | Text mining, machine learning, semantical analysis | Studies | Dictionary-based approaches, Deep Learning (Word2vec), Penalized Regression (elasticnet) | |
| Institut national de la statistique et des études économiques (INSEE) and Statistical office of the Interior Ministry | Identify the cases of intra-family violence in the complaint filings | The purpose of this study is to identify for each complaint, whether it can be considered as a intra-family violence or not, by exploiting contextual information filled out in a free-format box together with standardized information. The experiment uses 3 millions of complaints from the Paris region and applies textual analysis and machine learning techniques | Text mining, classification | Stand by | LDA, Random Forest | R |

---

[46] I. R. Kazimir, G. Horvath and J. Giczi, "Modeling Tax Evasion of the Self-employed in Hungary"

| Institution | Project name | Description of the project | Application | Status | Method | Software |
|---|---|---|---|---|---|---|
| Institut national de la statistique et des études économiques (INSEE) and IPP-Paris School of Economics | Machine learning for predicting careers and wages for micro-simulation models | Modelling labour market trajectories is of crucial importance to study retirement behaviours, pension distribution, and financial balances of retirement plans, especially in a pay-as-you-go system. We explore two margins of improvement for microsimulation of labour market trajectories: the use of non-parametric methods (random forests and boosted trees) and a discretized modelling of individual unobserved heterogeneity. We use administrative data from French complementary pension system in the private sector to design and test our simulation method; those data lack variables related to education level, marriage and family. Performance of models is assessed regarding both cross-sectional indicators and indicators of trajectory consistency. By doing so, we ensure that the techniques we develop reproduce the inner individual dynamics, a real challenge when simulating pension levels. In line with the needs expressed in reviews of existing models, we propose a simple framework for a systemic evaluation of models' performance. | Multinomial classification | Ongoing experiment | Classification Trees, Random Forest, Boosting | R |
| Institut national de la statistique et des études économiques (INSEE) | Use of machine learning algorithms for nonresponse issues through reweighting | The objective of this study is to improve the quality of nonresponse treatments in a general context of increasing nonresponse rates in both firm and household surveys. We want to explore the ability of machine learning algorithms to take the full advantage of the huge amount of auxiliary information coming from different horizons such as, paradata, administrative data, location information, etc, in a very flexible way. Exporatory works are run on the Adult education survey and the Structural Business Statistics Survey. | Supervised algorithms | Ongoing study | Random Forest, Bagging, Boosting, CART, Gradient Boosting Tree, Group Lasso. | R |
| Italian National Institute of Statistics | Substitutes for surveys via internet scraping[47] | Research regarding the possibility of substituting (fully or partially) surveys by collecting data via internet scraping and extracting information therein using machine learning methods. | | Experiment | Naive Bayes and others | R |
| National Institute of Statistics Romania | Use of administrative data in business statistics | Efficient integration of administrative data into the statistical process implies finding and resolving data quality issues. methods are used for imputation of the turnover variable for businesses from the value added tax (VAT) administrative data. | Data imputation | Experimental | Tree-Based Methods: Regression Trees, Random Forest, Boosting | R |

---

[47] Barcaroli, G., Nurra, A., Salamone, S., Scannapieco, M., Scarno, M., Summa, D. (2015). Internet as Data Source in the ISTAT Survey on ICT in Enterprises. Austrian Journal of Statistics, 44, 31–43, http://www.ajs.or.at/index.php/ajs/article/view/vol44-2-3/62 (last accessed 10 December 2018).
https://slideplayer.com/slide/7600964/ (last accessed 10 December 2018).

| Institution | Project name | Description of the project | Application | Status | Method | Software |
|---|---|---|---|---|---|---|
| National Institute of Statistics Romania | Price index estimation using web scraped data | Levenshtein distance between two strings is the number of deletions, insertions or substitutions required to transform source string into target string. This string matching technique is necessary for automatic classification of products names into categories and across periods. | String matching | Experimental | Levenshtein distance | R |
| National Institute of Statistics Romania | Action plan for EU-SILC improvements | The data matching methods have been used in order to increase the quality of EUSILC sample bringing together information from different data sources: sample survey and administrative registers. | Poverty indicators | Productive | Random Forest | R |
| National Institute of Statistics Romania | Modeling the potential human capital on the labor market[48] | Creating the profile of two categories of potential human capital by modelling the relationship between economically inactive persons who are seeking for a job, but are not immediately available to start working, respectively economically inactive persons who are not seeking for a job, but are immediately available to start working, and some socio-economic predictors. The aim is to identify the impediments which determine inactive people not to become active on the labour market. | Classification | Actually no ML; experiment | Logistic Regression | R |
| National Statistics Center, Japan | Supervised multiclass classifier for an autocoding for the Family Income and Expenditure survey[49] | Multiclass classifier that can classify Japanese short text descriptions according to their corresponding classification codes has been developed for the Family Income and Expenditure survey in Japan. The concept of the naïve Bayes classifier is borrowed for the algorithm of the classifier.<br>The classifier is also applicable to English text descriptions and other classifications tasks. | Multiple classification | Experiment | Naïve Bayes | Pearl / R |

---

[48] Ciuhu, A.-M., Caragea, N., Alexandru, C. (2017). Modeling the potential human capital on the labor market using logistic regression in R. Romanian Statistical Review, 4, 141–152, http://www.revistadestatistica.ro/wp-content/uploads/2017/11/RRS-4_2017_tipar_A11.pdf (last accessed 10 December 2018).
[49] http://www.nstac.go.jp/services/society_paper/29_02_01_poster.pdf (last accessed 10 December 2018).
 Toko, Y., Wada, K., Kawano, M. (2017). A Supervised Multiclass Classifi er for an Autocoding System. Romanian Statistical Review, 4/2017, 29–39, http://www.revistadestatistica.ro/wp-content/uploads/2017/11/RRS-4_2017_A02.pdf (last accessed 10 December 2018).

| Institution | Project name | Description of the project | Application | Status | Method | Software |
|---|---|---|---|---|---|---|
| OECD | Algorithms and Colllusion[50] | The combination of data with technologically advanced tools such as pricing algorithms and machine learning is increasingly changing the competitive landscape in the digital markets. There is a growing number of firms using computer algorithms to improve their pricing models, customise services and predict market trends, which could generate efficiencies. However, the widespread usage of algorithms could also pose possible anti-competitive effects by making it easier for firms to achieve and sustain collusion without any formal agreement or human interaction. | Classification | Experiment | Several methods | |
| ONS | Unsupervised document clustering with cluster topic identification[51] | This piece of work is inspired by a project the Office for National Statistics' (ONS's) Big Data Team have been exploring with data extracted from Companies House. Medium- to large-sized businesses are required to submit a full accounts document to Companies House and within this document the businesses typically include a description of the function of the company. This is potentially useful information to help classify the business into its relevant UK Standard Industrial Classification 2007: UK SIC 2007 code, and may offer insight into emerging new topics in industry. | Clustering | Experiment | Doc2vec | |
| STATEC (Luxembourg) | Scanner Data | Machine Learning Ranking (MLR) is used in the Scanner Data project to classify the individual items into COICOP (Classification of individual consumption according to purpose), which is the standard classification used for compiling a CPI. It allows to use a larger sets of incoming data without increasing costs of manuel processing. | Classification | Productive | MLR | Solr / Java / Talend |

---

[50] OECD (2017). Algorithms and Collusion: Competition Policy in the Digital Age. Report, http://www.oecd.org/daf/competition/Algorithms-and-colllusion-competition-policy-in-the-digital-age.pdf (last accessed 10 December 2018).
[51] https://www.ons.gov.uk/methodology/methodologicalpublications/generalmethodology/onsworkingpaperseries/onsworkingpaperseriesnumber14unsuperviseddocumentclusteringwithclustertopicidentification (last accessed 10 December 2018).

| Institution | Project name | Description of the project | Application | Status | Method | Software |
|---|---|---|---|---|---|---|
| STATEC (Luxembourg) | Business Enterprise Research and Development (BERD) | An in-house developed ensemble classifier assesses for each survey respondent the probability of performing intra-mural R&D activities during the reference year. The probability is conditional on the available data, which includes current and past survey data (including unstructured text) as well as administrative sources. The model results are used in the survey data validation process to spot item non-response and inaccurate responses on the intra-mural R&D variables. Such respondents are then individually contacted by the statistical analysts, if necessary. Albeit the use of an ensemble classifier, the model results remain fully interpretable. | Classification | Productive | Model Stacking | KNIME, R |
| Statistics Austria | Imputation[52] | In various surveys and also in projects with administrative data missing values should be imputed. | Imputation | Productive | kNN | R |
| Statistics Austria | Statistical matching | On the basis of common variables two data sets are matched (very similar to imputation.) | Statistical matching | Productive | Random Forest, kNN | R |
| Statistics Austria | Estimating AROPE for the Austrian rich frame[53] |  | Estimation/Classification | Productive | Boosting Trees Algorithm | R |
| Statistics Austria | Estimating a SILC-like income based on adminstrative data | For a couple of years now, the household income in the ICT survey is not asked for bute estimated based on adminstrative data and the SILC data. | Estimation/Regression | Productive | Random Forest | R |
| Statistics Belgium | Predicting the NACE code of a job vacancy | Machine learning is used to predict the NACE code of a job vacancy based on the job description. We use administrative databases (from national jobs portals) to model the link between words in the description and the NACE code. We expect to apply this model to scrap job vacancies from internet job portals in Belgium. | Binary classification | Test | SVM, | R (RtxtTools) |

---

[52] Kowarik, A., Templ, M. (2016). Imputation with the R Package VIM. Journal of Statistical Software, 74(7), https://www.jstatsoft.org/article/view/v074i07 (last accessed 10 December 2018).
[53] Alfaro, E., Gamez, M., García, N. (2013). adabag: An R Package for Classification with Boosting and Bagging. Journal of Statistical Software, 54(2), http://www.jstatsoft.org/v54/i02/ (last accessed 10 December 2018).

| Institution | Project name | Description of the project | Application | Status | Method | Software |
|---|---|---|---|---|---|---|
| Statistics Canada | Consumer Prices scanner data use | Consumer Prices: retail scanner data classification to the Consumer Price Index (CPI) commodities classification, used to suggest CPI product substitutions. Currently, in production parallel run. | Classification | In production parallel run | SVM | python |
| Statistics Canada | Retail scanner data use for Monthly retail trade survey and Quarterly retail commodity survey | Machine learning text classification is used to obtain the NAPCS code of each product sold within the retail scanner data, and obtain aggregate sales for each NAPCS, aggregate sales by area/ postal code. Proof of Concept (PoC) completed. | Classification | Transitioning to production | XGBoost linear, with bag of words character n-grams model | R |
| Statistics Canada | International Trade data Outliers Detection | Outliers are a major problem in the international trade data, in particular for the quantity variable. Errors include unit errors, 0 or 1 entered instead of a proper quantity,etc. Current system is based on unit value (UV) clipping, manual checking (including "unclipping"), and an approval process. Machine learning (ML) is used to automate this process. ML is also used to reconstruct/ impute the original value of the flagged points. | Outlier detection | The outlier detection experiment is completed. The imputation work is in progress. | XGBoost Tree Model | R |
| Statistics Canada | Business Activity, Expenditure and Output (BAEO) survey comments text mining | The BAEO survey receives close to 9000 comments. ML was used to classify the comments into 8 action categories. | Classification | Experiment completed | Linear SVM, with bag of words model | R |
| Statistics Canada | Payments data feasibility project | ML is used to classify the payment transactions into standard statistical classification concepts (NAICS, COICOP, etc). The project's goal is to determine if the data can be used to produce information related to retail sales, household consumption, digital transactions, and tourism statistics.<br>ML might be also used for imputation of misisng values. | Classification, imputation | At experimentation stage | SVM | R |

| Institution | Project name | Description of the project | Application | Status | Method | Software |
|---|---|---|---|---|---|---|
| Statistics Canada | Transport statistics: Trucking Commodity Origin and Destination Survey (TCOD) | ML will be used to classify the electronically reported data. In the context of the TCOD survey redesign, the use of auxiliary sources such as GPS data, satellite imagery is evaluated as a new source for trucking data analysis, including linkage to a specific business on the BR. | Text classification, image classification, predictive modelling for missing information based on auxiliary data, record linkage | At experimentation stage | recently started | R, Python |
| Statistics Canada | Enterprise statistics | Web scraping for large enterprises to complement the survey data sources with on-line data (newsfeed, financial reports, company Web sites) to enhance data coherence, improve profiling, prepare the Entreprise Portfolio Managers' visits to companies. | Data extraction, Natural Language Processing, Record Linkage | At experimentation stage | recently started | R, Python |
| Statistics Canada | Web scraping to enhance the research and development survey frames | Web scraping is used to supplement the R&D survey data with on-line data to identify adequately the survey population, and develop complementary indicators for innovation. | Data extraction, Natural Language Processing, Record Linkage | At experimentation stage | recently started | R, Python |
| Statistics Canada | Retail statistics | Web scraping and google map for retail store information | Data extraction, Natural Language Processing, Record Linkage | At experimentation stage | Web scraping libraries | Python |
| Statistics Canada | Manufacturing industries and producer prices commodities | Web scraping to obtain the list of commodities being produced by manufacturing companies from catalogues on their Web sites or other existing on-line data bases. Code the products to the NAPCS classification using ML. | Data extraction, Classification to NAPCS | At test stage, experimentation to start in May'18 | TBD | R, Python |

Martin Beck; Florian Dumpert; Joerg Feuerhake: Machine Learning in Official Statistics| Institution | Project name | Description of the project | Application | Status | Method | Software |
|---|---|---|---|---|---|---|
| Statistics Canada | Agriculture crop yield estimates[54] | Phase 1 : The model-based crop estimates provide provincial and national yield and production estimates for principal field crops in Canada. The model utilizes data from low resolution satellite imagery, historical field crop survey estimates, and agroclimatic information.<br>Phase 2: Develop in season crop area and yield estimates, combining crop insurance, remote sensing and other business intelligence including supply and disposition data and historical patterns to identify likely crop cover at the field level and then employing historical data, trends and weather data to produce yield estimates. | Predictive modelling | Phase 1 in production since 2015<br>Phase 2 (November crop yield and area) at Idea stage, to start in May'18 | Phase 1 : LASSO and others<br>Phase 2 : TBD | mostly SAS for phase 1; TBD for phase 2 |
| Statistics Canada | Canadian Housing Statistics Program, project on citizenship and country of birth | Partial information about citizenship and country of birth can be found in different databases such as the Census of Population, the Social Insurance Number Registry, immigration data, tax data, and so on. Standardize and integrate the existing information, use ML to impute missing information. | Classification, imputation | Idea, to start in May'18 | TBD | TBD |
| Statistics Canada | Exploration of machine learning to create indicators on tourism spending | Model a set of early indicators on international tourism spending in Canada based on survey data and payment processor data from debit and credit cards. | Regression | Idea | TBD | TBD |
| Statistics Canada | Economic Analysis: Identifying high growth firms and firm failures | Administrative data over a three-year time span is commonly used to identify high growth firms. This project uses ML techniques to develop more timely estimates of high growth firms. Similarly, firm peformance is hypothesized to deteriorate years before exit (shadow of death effect).   ML techniques are also applied to develop more timely estimates of firm exits. | Classification | Idea, to start in May 2018 | Random Forest | R |

---

[54] http://www23.statcan.gc.ca/imdb/p2SV.pl?Function=getSurvey&SDDS=5225 (last accessed 10 December 2018).

| Institution | Project name | Description of the project | Application | Status | Method | Software |
|---|---|---|---|---|---|---|
| Statistics Canada | Economic Analysis: Identifying firm networks | Using ML to identify firm networks from Business Register and origin and destination of shipments data so that supply chains and intra- and intra-firm trade can be analysed | Classification/Network Analysis | Initial assessment completed. Additional data needed to improve algorithm. | TBD | TBD |
| Statistics Canada | Communications and Dissemi-nation: improving user experience on the web | Live chat featured for visitors based on their navigation patterns | Chatbots | Idea | TBD | TBD |
| Statistics Canada | Communi-cations and Dissemi-nation: improving user experience on the web | Use AI to analyze content web site visitors based on their past content consumption / navigation, or to offer live chat. | TBD | Idea, to start in April 2018 | TBD | TBD |
| Statistics Canada | Predicting Mortality Rates | Investigate how sociodemographic and health-related factors contribute to life expectancy. Constructing well-specified predictive models can be extremely time consuming and labor intensive. Data driven approaches, like machine learning techniques, are gaining popularity. These algorithms may be able to gain insightful information about what predicts an outcome by iteratively learning from data, instead of being explicitly directed by theory. ML solution: Trial various AI/machine learning techniques, including support vector machines and neural networks, to predict mortality in the Census linked CANCHEC cohorts. These are large cohorts that contain health outcomes, but not necessarily health exposure data. As Statistics Canadas holdings of such data increase we need to explore alternative methods to construct robust predictive models/analytic tools, when we are lacking key health exposure variables. | Predictive modelling | Idea to start in FY 2018 | TBD | TBD |

| Institution | Project name | Description of the project | Application | Status | Method | Software |
|---|---|---|---|---|---|---|
| Statistics Canada | Harvesting key data on causes of death from narrative descriptions | Harvesting key data on causes of death and abuse from narrative descriptions Exploring narrative descriptions to harvest statistical information (cause of death in coroners' reports; abuse case in social workers' narrative description)<br>A machine learning application is being used to mine more rapidly and efficiently the unstructured narratives that coroners include in their reports and that detail the circumstances of the deaths. These narratives are included in our Canadian Coroners and Medical Examiners Database (CCMED). Its first goal is to identify opioid-related deaths. In the longer-term, it is hoped that machine learning/Artificial Intelligence would permit to rapidly recognize any patterns that would point to another crisis or specific circumstances that affect many investigated death cases. | Classification | Exploration part 1 completed (promising approach with limited success due to weak training data); Literature review completed for future exploration | Various (SVM, NN, Adabost Naïve Bayes) | Python |
| Statistics Canada | Victimisation studies (Harvesting key data from narrative descriptions) | Exploring narrative descriptions to harvest statistical information (abuse case in social workers' narrative description) | Classification of narrative description | Idea, to start in Spring/Summer 2018 (data dependent) | Text recognition (possibly), Natural language processing, TBD | TBD |
| Statistics Canada | R&D for generalised systems[55] | Use of genetic algorithm (AI based on natural selection) for sample allocation | Sample allocation/selection | Proof of concept completed | Genetic Algorithm | R |
| Statistics Canada | Census | Immigration admission category variables were added to the 2016 Census through record linkage rather than collecting them from respondents.<br>Data were not available for some individuals and had to be inferred from other characteristics provided by them. Machine Learning was used to identify the best combination of characteristics to make these inferences. | Imputation | Completed and used for Census 2016 production | ReliefF algorithm for feature selection and weigthing for neirest neighbor imputation | R for feature selection and weigthing; model used in "regular" production system ( CANCEIS) |

---

[55] Ballin, M., Barcaroli, G. (2013). Joint determination of optimal stratification and sample allocation using genetic algorithm. Survey Methodology, 39(2), 369–393, http://www.statcan.gc.ca/pub/12-001-x/2013002/article/11884-eng.htm (last accessed 10 December 2018).

| Institution | Project name | Description of the project | Application | Status | Method | Software |
|---|---|---|---|---|---|---|
| Statistics Canada | Study of comments submitted during the 2016 Census content consultation | Used text mining software to analyze over 1.1 million comments compiled from the 2016 Census content consultation to inform possible content and questionnaire changes for 2021. | Key term identification and binary classification | Productive | Natural Language Processing | SAS JMP |
| Statistics Canada | Census and others | Exploring information provided in comments box in the Census with focus on non-binary gender self-identification in open text/comment box | Classification | Exploration | Unsupervised ML to cluster comments | SAS Enterprise Miner |
| Statistics Canada | R&D for imputation method | Hackathon for advanced imputation methods using AI. Use use is imputation of a large admin files where adata can be entered for generic or detailed financial item. When generic is used, data must be imputed for detailed item. | Imputation | Exploration via hackathon planned in May 2018 | Various | Various |
| Statistics Canada | Exploration of machine learning for coding of industry and occupation text descriptions | Opportunities to use Machine Learning and Artifical Intelligence to improve the effectiveness of automated coding of survey responses is being investigated. This would have a number of applications in survey and administrative data programs, including the Labour Force Survey. For example, new methods could permit the collection of additional open-ended information related to task descriptions of jobs and skills profiles of individuals. | Statistical coding | Exploratory / idea generation | Various | Various |
| Statistics Canada | Census Program Transfor-mation Project | Exploring how AI could be applied in data linkage processes in the building of statistical registers. | To be defined | Idea | TBD | TBD |
| Statistics Canada | Creating synthetic data for micro-simulation | Microsimulation models require complete data. Currently in the Population Health Microsimulation Model (POHEM) we do not model measured data such as that collected Canadian Health Measures Survey (CHMS). Having measured data in POHEM is a desirable attribute if the model is to be used for evidence-based policy analysis around cardiovascular disease and other chronic diseases. AI solution: Use AI/machine learning techniques to match data from the CHMS to the Canadian Community Health Survey (CCHS) in order to initialize a starting POHEM population with measured health data. | Classification, matching | Idea to start in FY 2018 | Likely K Nearest Neighbours | R, knnnn package |

| Institution | Project name | Description of the project | Application | Status | Method | Software |
|---|---|---|---|---|---|---|
| Statistics Canada | Synthetic Data File | Examination of machine learning algorithms for creation of synthetic data for disclosure control | Synthetic data (disclosure control) | Exploration | CART models and Random Forest | R |
| Statistics Canada | Priorisation score for collection | Creating a Priorization Score in CATI Surveys: Analysis of the Bayesian Hierarchical Rule Modeling | | Exploration / proof of concept in progress | Bayesian Hierarchical | SAS |
| Statistics Canada | Simulation of administrative data | Modeling of administrative data which can incorporate both the relationships between variables (correlations) and the longitudinal dependences are requirements to provide realistic simulated data for statistical purposes. Current methodology and available tools apply Gibbs Sampling or Bayesian networks to model all the conditional distributions. Machine learning algorithms may also provide a solution. For example, neural networks are capable of modeling complex correlations amongst variables. Such networks could be extended to deep networks in the case of large, complicated data sets. | Modelling | Idea | TBD | TBD |
| Statistics Canada | Automated extraction of features for record-linkage | The goal is to automate the selection of features for record-linkage. Currently these features are selected manually based on expert knowledge. | Classification, feature selection | Initial exploration | Supervised, unsupervised, dimension reduction techniques (e.g. PCA) | SAS |
| Statistics Canada | Development of the record linkage software G-Link | Supervised and Semi-supervised Machine Learning methods are used for the development of automatic thresholds in the Felleigi-Sunter methodology. The k-means method and the two-step method (k-means and probit) will be implemented in the version 3.4 of G-Link. In this way, Statistics Canada is creating an opportunity to replace costly searches carried out by humans through manual review by machine learning techniques. | Probabilistic record linkage with Fellegi-Sunter methodology | Development | Unsupervised: K-Means Supervised: two-step method K-means + Probit | G-Link (coded in SAS) |
| Statistics Canada | Machine Learning for record linkage | Similarities metrics such as Jaro, Jaro-Winkler, Fuzzy Winkler, Fuzzy Jaro are developed in G-Link and ready to be used by Machine Learning methods outside of G-Link. We are exploring the use of the SAS enterprise Miner software.<br>In this way, Statistics Canada is creating an opportunity to examine and compare the Fellegi-Sunter classifier with alternative methods (SVM, neural networks etc. ) | Binary classification | Exploration | SVM, Neural network, Supervised Logistic | SAS EM |

| Institution | Project name | Description of the project | Application | Status | Method | Software |
|---|---|---|---|---|---|---|
| Statistics Canada | NAICS/NOC autocoder | Automatic data classification (various classifications including occupation and industry) | Classification to NAICS/NOC | Refining the model for increased accuracy, to be implemented in Python | Bag of words and statistical classifier | Pearl, Python |
| Statistics Canada | Exploration of machine learning to identify driving under the influence by type of substance | Social media scrapping for estimating the prevalence of driving under the influence by type of substance. | Classification | Idea | TBD | TBD |
| Statistics Denmark | Imputation of educational status for immigrants | | Imputation | | Random Forest | R (missForest) |
| Statistics Finland | Machine reading accident reports | Free-text road traffic accident reports from the Police are classified to those which caused personal injuries and those which did not | Binary classification | Productive | tf-idf + Logistic Regression | Python |
| Statistics Finland | Automatic coding of industry and occupation[56] | Random forest classifiers are used to automatically classify Finnish Labour Force Survey respondents to correct industry and occupation (NACE, ISCO) based on combined register and survey data | Multi-class classification | Development | Random Forest | Python |
| Statistics Iceland | Assignment of fine-grained product ids based on product descriptions | | Classification | In review | Random Forest | Python |
| Statistics Iceland | Increased automation in data processing | | Varied | Planning | N/A | N/A |

---

[56] https://www.bigsurv18.org/program2018?day=3 (last accessed 10 December 2018).

Martin Beck; Florian Dumpert; Joerg Feuerhake: Machine Learning in Official Statistics| Institution | Project name | Description of the project | Application | Status | Method | Software |
|---|---|---|---|---|---|---|
| Statistics Netherlands | | Clustering and classification of SMEs (small and medium enterprises) based on website descriptions | Clustering | Research | k-Means | |
| Statistics Netherlands | | Imputation of economic data | Imputation | Research | Random Forest | |
| Statistics Netherlands | Inference from non-probability samples | Different methods are compared to make inference from simulated samples generated through mechanisms other than random sampling to correct for selection bias. Applied to data on annual mileages of vehicles from the Dutch Online Kilometer Registration. | Regression | Experiment | Nearest Neighbor, Neural Network, Regression Tree, Support Vector Machine | R |
| Statistics Netherlands | Predicting moving behavior | Random forests are used to predict moving behavior of persons from registered life-history events. Predictions can potentially replace a survey questionnaire and thus contribute to shorter questionnaires and reduce survey fatigue. | Binary classification, can be expanded to multinomial classification | Experiment | Random Forest | R |
| Statistics Netherlands | Improvement of imputation by robust estimation and machine learning methods[57] | Standard automatic imputation methods for business statistics, such as ratio-imputation are not always accurate. We are looking for improvements by using robust estimators and automated model building by machine learning techniques (gradient boosting machine). | Imputation of business statistics | Test | GBM (Gradient Boosting Machine) | |
| Statistics Netherlands | URL retrieval Enterprises[58] | Retrieval of URLs using Google custom search, a training set of enterprises with known URLs and a list of enterprises with administrative infomation. For Eommerce detection, social media detection and identifying foreign companies. | Binary classification | Production | Decision Tree / Random Forest | Node + Python |
| Statistics Netherlands | Enterprise classifier[21] | Prediction of website use, ecommerce detection, social media detection. | Binary and multiclass classifiers | Test | Decision Tree / Random Forest | Node + Python |

---

[57] Pannekoek, J. (2018). Improvements of ratio-imputation using robust statistics and machine learning-techniques. Working Paper for the UNECE Workshop on Statistical Data Editing September 2018, https://www.unece.org/fileadmin/DAM/stats/documents/ece/ces/ge.44/2018/T6_Netherlands_PANNEKOEK_Paper.pdf (last accessed 10 December 2018).
[58] Stateva, G., ten Bosch, O., Windmeijer, D., Maslankowski, J., Barcaroli, G., Scannapieco, M., Greenaway, M., Jansson, I., Wu, D. (2018). ESSnet Big Data Specific Grant Agreement No 1 (SGA-2), Work Package 2 Web scraping Enterprise Characteristics, Final Report, https://webgate.ec.europa.eu/fpfis/mwikis/essnetbigdata/images/e/ee/Wp2_Del2_4.pdf (last accessed 10 December 2018).

| Institution | Project name | Description of the project | Application | Status | Method | Software |
|---|---|---|---|---|---|---|
| Statistics Netherlands | Classification of products into product categories for the CPI | The Random Forest algorithm has been extended with active and online learning techniques to classify online products (scraped from e.g. webshops) into product categories. The algorithm aims to keep the training set as small as possible by re-using training items from previous months. Only when quality gets too low, additional training items from the current month are added to the training set, and we then try to use those items that result in the biggest improvement in quality. | Multiclass classification | Idea | Random Forest | Python |
| Statistics Netherlands, in collaboration with Tenforce | Classification of supermarket products for the CPI | Research into the use of Support Vector Machines and Random Forest to classify supermarket (food) items into LCOICOP categories using scanner data | Multiclass classification | Test | SVM / Random Forest | Python |
| Statistics Netherlands | Domain names ICT companies | A nation wide URL domain name data base (the .nl domain) is connected to the outcome of the Dutch ICT-company survey in order to create a ML-trainingset based on the web-scraped contents of the company's URL. In this way, ICT-company classification can be automated and the ICT-company statistics can be improved | | Experiment | | Python |

| Institution | Project name | Description of the project | Application | Status | Method | Software |
|---|---|---|---|---|---|---|
| Statistics Netherlands | Adaptive data collection at Statistics Netherlands with an application to the Health Survey[59] | Challenges that surveys are facing are increasing data collection costs and declining budgets. During the past years, many surveys at Statistics Netherlands were redesigned to reduce cost and to increase or maintain response rates. Currently, alternative approaches are investigated to produce more accurate estimates within the same budget. Adaptive data collection is proposed for achieving this goal. Research into the effect of reducing face to face observation in mixed mode surveys on quality and costs was carried out in 2017. Reducing face to face observation can be done in various ways. It can be done through random selection, but also through stratified selection of nonrespondents eligible for face to face follow-up. By using the latter method, nonresponse bias can potentially be reduced. The key decisions to be made are how to divide the population into strata and how to compute the allocation probabilities for face to face follow-up in the different strata. In this presentation the adaptive data collection is elaborated for the Health Survey as it is conducted by Statistics Netherlands since 2018. Attention is paid to the choice of the strata, the choice of the mixed mode observation strategy, the optimization problem with corresponding constraints and the effect of the adaptive data collection on most important survey estimates. | Classification | Productive | Classification Tree | R |
| Statistics Netherlands | Prediction of economic activity from website texts[60] | Economic activity is an important characteristic of enterprises. This characteristic is stored in a General Business Register at CBS, but it tends to be outdated. We aim to use machine learning techniques to predict the economic activity from website texts | Multiple cases | Test | Various ML techniques like SVM and Random Forest | R and Python |
| Statistics Netherlands | Cyber Security | Predict the number of cyber security companies in The Hague area, based on the company websites. We use webscraping, textming and machine learning techniques. | Binary classification | Experiment | SVM | Python |

---

[59] van Berkel, K. (2018). Adaptive data collection at Statistics Netherlands with an application to the Health Survey. Working Paper for the UNECE Workshop on Statistical Data Collection October 2018, http://www.unece.org/fileadmin/DAM/stats/documents/ece/ces/ge.58/2018/mtg7/DC2018_S5_Berkel_Netherlands_AD.pdf (last accessed 10 December 2018).
[60] Roelands, M., van Delden, A., Windmeijer, D. (2017). Classifying businesses by economic activity using web-based text mining. Discussion Paper, https://www.cbs.nl/en-gb/background/2017/47/classifying-businesses-by-economic-activity (last accessed 10 December 2018).

| Institution | Project name | Description of the project | Application | Status | Method | Software |
|---|---|---|---|---|---|---|
| Statistics Netherlands | Automated classification of sustainable companies based on machine learning | Predict the number of companies active in the field of specific SDGs based on the company websites. We use webscraping, textming and machine learning techniques. | Binary classification | Experiment | Yet unknown | Python |
| Statistics Netherlands | Deep learning - Estimating missing values | Turnover data of companies contain a lot of missing values. Imputation techniques may be improved by using machine learning techniques. The aim of this study is to estimate missing values by using deep learning. | Prediction | Experiment | Deep Learning (Neural Networks) | Python |
| Statistics Netherlands | Cross-Border Internet Purchases at EU webshops[61] | Semi-automated machine learning to predict whether an EU company sells goods to Dutch consumers through webshops. | Binary classifcation | Productive (beta) | Semi-Automated Machine Learning | Python |

---

[61] Meertens, Q. A., Diks, C. G. H., van den Herik, H. J., Takes, F. W. (2018). A Data-Driven Supply-Side Approach for Measuring Cross-Border Internet Purchases. Cornell University arXiv:1805.06930, https://arxiv.org/abs/1805.06930 (last accessed 10 December 2018).

| Institution | Project name | Description of the project | Application | Status | Method | Software |
|---|---|---|---|---|---|---|
| Statistics Norway | Exploration on Random Forest for editing purposes in register based salary statistics | | Classification | | Random Forest | R |
| Statistics Poland | Detection of agricultural crops[62] | The goal is to identify crop types based on Sentinel-1 and Sentinel-2 satellite images. Different algorithms have been tested for the detection of crop types, including Support Vector Machine (SVM), Decision Trees (DT), K-Nearest Neighbours (KNN) with the following classification parameters: Sigma, Entropy, Alfa, multi-temporal indicators, Wishard distribution but the highest accuraccy is based on KNN with wishard distribution. | Segmentation | Pilot | KNN, SVM | MTSar, ArcGIS |
| Statistics Poland | Life satisfaction | The goal of the use case is to deliver data on life satisfaction - 1.happy, 2.neutral, 3.calm, 4.upset, 5.depressed and 6.discouraged. The goal is to support the data from EU-SILC survey with more recent data. The major drawback from this case study is that the dataset may not be representative. The methodology includes Machine learning – supervised learning and Web scraping – we use Twitter API to gather and process the data. | Classification | Pilot | NL | Python, MongoDB, Apache Spark |
| Statistics Portugal | Big Data ESSNet | | Multiclass classification | Test | SVM (linear kernel) | Python (SciKit-Learn library in a Jupyter Notebook environment) |
| Statistics Portugal | Big Data ESSNet | | Multiclass classification | Test | Perceptron (linear) | Python (SciKit-Learn library in a Jupyter Notebook environment) |

---

[62] WP 7 team (2018). ESSnet Big Data Specific Grant Agreement No 1 (SGA-1), Work Package 7 Multi domains, Report, https://webgate.ec.europa.eu/fpfis/mwikis/essnetbigdata/images/1/15/WP7_Deliverable_7.1_7.2_7.3_2017_02_01.pdf (last accessed 10 December 2018).

| Institution | Project name | Description of the project | Application | Status | Method | Software |
|---|---|---|---|---|---|---|
| Statistics Portugal | Big Data ESSNet | | Multiclass classification | Test | Neural Network Model for language identification | R Package "cld3" (Google's Compact Language Detector 3) |
| Statistics Portugal | Identification of error-containing records via classification trees[63] | A method based on classification trees for error detection in foreign trade transaction data collected by the Portuguese Institute of Statistics. | Error detection | Experiment | Decision Tree | |
| Statistics Spain (INE) | UFAES (new methodology) | Random forests are used to model questionnary data from admin data. This model assists the design-based estimator in a probability sampling survey in order to reduce the sample size. | Algorithm-assisted survey sampling | Research with real data | Random Forest | R |
| Statistics Spain (INE) | Selective Editing of Quantitative Variables | Different predictive techniques are to be explored to predict anticipated values in the optimization approach to selective editing of quantitative variables developed at Statistics Spain (INE) | Prediction | Planning stage | Random Forest, SVMs, Spline Regression, kNN Regression | R |
| Statistics Spain (INE) | Selective Editing of Qualitative Variables | Different classification techniques are to be explored to classify influential/non-influential units in the optimization approach to selective editing of qualitative variables under development at Statistics Spain (INE) | Binary classification | Planning stage | Random Forest, SVMs, Logistic Regression, kNN Regression | R |
| Statistics Sweden | Essnet big data WP2 - Web scraping enterprise characteristics - Use case of Job advertisement | | Binary classification | Under development | SVC, Decision Tree, Naive Bayes, Keras sequential NN | Python |

---

[63] Soares, C., Brazdil, P., Pinto, C. (2002). Machine learning and statistics to detect errors in forms: competition or cooperation. Proceedings of the ECML/PKDD Vol. 2, 19–23.

| Institution | Project name | Description of the project | Application | Status | Method | Software |
|---|---|---|---|---|---|---|
| Statistics Sweden | Essnet big data WP2 - Web scraping enterprise characteristics - Use case of NACE | | Multilabel classification | Under development | Keras sequential NN | Python |
| Statistics Sweden | Automatic coding of occupation title using machine learning methods | | Binary classification | Implementation in production | k Nearest Neigbour | C# and R |
| Stats NZ | Assignment of geographic region to individuals | Decision Trees and random forest methods used to assign the correct Territorial Authority to individuals given a range of administrative data sources | Classification | Test | Decision Tree | R |
| Stats NZ | Classification of Building Consents with Natural Language Processing | A generalised linear model has been trained on historical consents data to perform classification of building consents into multiple classes. The model uses a bag of words approach, and is currently being brought into production. | Multi-level classification | Tested, not yet in production | Supervised Learning: Generalised Linear Model | R |
| Stats NZ | Classification of building consents | We used manually coded data to train a model that categorises building consents based on a free text field (job description, as supplied by the applicant). We predict several variables, including building type (which has 29 possible values). | Classification | Experiment | Generalised Linear Model using vectorised n-grams | R (glmnet and text2vec) |
| Stats NZ | Coding of industry classification | Early days in thinking | Blassification | Idea | | |
| Stats NZ | Automatic coding of census variables via Support Vector Machines | Investigation of the potential of using Support Vector Machines (SVM) to improve coding of item responses in their Census. They applied SVM to code the variables Occupation and Post-school Qualification, using two disjoint sets of observations, each of size 10,000, from Census 2013 data for training and testing. | Multi-class classification | Experiment | SVM | |

| Institution | Project name | Description of the project | Application | Status | Method | Software |
|---|---|---|---|---|---|---|
| Stats NZ | Imputation via Classification and Regression Trees | Investigation in the use of CART to predict two binary variables based on Census 2013 data. The binary variables were<br>1. the missingness of the income variable, and<br>2. the response to the question of whether the respondent has moved since the previous census. Results of this investigation are being evaluated. | Classification / imputation | Experiment | Decision Tree | |
| Stats NZ | Determination of imputation matching variables | Statistics New Zealand is redesigning the editing and imputation methodology of their Household Economic Survey (HES). Their current proposed methodology will use the Canadian Census Edit and Imputation System (CANCEIS). The imputation module of CANCEIS is based on the Nearest Neighbour Imputation Methodology, which requires user specification of a distance measure of pairs of units based on a number of "matching variables" as well as weights which defines the relative importance of these matching variables. The weight of a matching variable should reflect its strength as a predictor for the variables to be imputed. Statistics New Zealand has reported promising results in using Random Forests to select the set of matching variables for CANCEIS, as well as their weights. | Imputation / variable selection | Experiment | Random Forest | |
| Stats NZ | Creation of homogeneous imputation classes | Comparison of two methods (CART, predictive mean stratification) for creating homogeneous imputation classes. | Imputation | Experiment | Decision Tree | R |
| Stats NZ | Derivation of edit rules | Investigation of the potential of using association analysis to derive additional edit rules to enhance the processing of census data. | Editing | Experiment | Association Analysis | |
| U.S. Bureau of Economic Analysis | Nowcasting of Source Data for Advance GDP Estimates | This pilot effort aims to reduce revisions in key GDP components by improving trending methods. By training an ensemble of ML models (e.g. LASSO, Ridge, Random Forest, etc) using a variety of alternative data, nowcasted predictions are produced for certain source data series in time for the advance estimate of national GDP. Note that this project is currently being evaluated in parallel with the current estimate process. | Regression | Pilot | Ensemble Methods | R |

| Institution | Project name | Description of the project | Application | Status | Method | Software |
|---|---|---|---|---|---|---|
| U.S. Bureau of Labor Statistics | Automatic coding of worker injury narratives for the Survey of Occupational Injuries and Illnesses[64] | Each year the Survey of Occupational Injuries and Illnesses (SOII) collects hundreds of thousands of written narratives describing work related injuries and illnesses. In order to produce statistics from this information each of these narratives receives 6 of several thousand possible codes to indicate the occupation of the worker and various characteristics of the incident. To improve the consistency and efficiency of this manual effort BLS developed and evaluated a variety of automated approaches, settling initially on regularized multinomial logistic regression. BLS found automated coding with regularized logistic regression produced more accurate coding than trained human coders, even after the human codes had the benefit of several layers of review. As a result BLS began using this technique to automatically assign codes starting with 2014 data. BLS is now automatically assigning nearly two-thirds of all SOII codes and has recently developed new deep neural network models that provide even better performance. These neural network models are currently used to identify suspicious codes for review, and are likely to be deployed for production autocoding later this year. | Multinomial text classification | Production | Regularized Logistic Regression, Deep Neural Networks | Python, Scikit-learn, Tensorflow, Keras |
| U.S. Bureau of Labor Statistics | Automatic coding and review of occupation narratives for the Occupational Requirements Survey | One of the key data elements collected by the Occupational Requirements Survey (ORS) is the occupation classification, which is recorded in part, in text format. To more efficiently and effectively validate this data BLS is investigating the application of the same machine learning techniques now successfully being used for the SOII to assist in the review and validation of ORS occupation data. Initial results show promise. BLS will also be investigating whether occupation data from other surveys (like the SOII, and the National Compensation Survey), can be used to improve the performance of the ORS automatic reviewing system. | Multinomial text classification | Research | Regularized Logistic Regression | Python, Scikit-learn |

---

[64] https://www.bls.gov/iif/autocoding.htm (last accessed 10 December 2018).
Measure, A. C. (2014). Automated coding of worker injury narratives. JSM, https://www.bls.gov/osmr/pdf/st140040.pdf (last accessed 10 December 2018).
Measure, A. C. (2017). Deep neural networks for worker injury autocoding. Discussion Paper, https://www.bls.gov/iif/deep-neural-networks.pdf (last accessed 10 December 2018).

| Institution | Project name | Description of the project | Application | Status | Method | Software |
|---|---|---|---|---|---|---|
| U.S. Bureau of Labor Statistics | Automatic extraction of benefits information from Summary of Benefits and Coverage documents. | Health insurance benefits are an important component of worker's compensation and are often described in a semi-structured document called the "Summary of Benefits and Coverage". This project aims to use machine learning to automatically extract benefits information from these documents with the goal of eventually using this information to augment the National Compensation Survey. Initial results demonstrate that we can automatically extract some of this information at very high accuracy. | Text classification, information extraction | Research | Random Forest | Python, Scikit-learn |
| U.S. Bureau of Labor Statistics | Automatic linkage of fatal injury case information to OSHA records | To produce statistics about fatal occupational injuries in the U.S., the Census of Fatal Occupational Injuries collects and combines information from a wide variety of sources including local, state, and federal government agencies and U.S. media. One of the biggest sources of official information is data from the Occupational Safety and Health Administration, which often investigates fatal work related injuries. One of the key challenges in incorporating this information is figuring out whether an OSHA record corresponds to a record already in the master file, and if so, which one. Often, this must be accomplished even without imperfect identifiers like decedent and establishment name. To address this issue this project uses machine learning, trained on previously linked documents, to automatically determine whether an OSHA investigation document should be linked to an already partially collected case, or represents a new case that should be added to the master file. By combining a variety of noisy and sometimes missing signals including information about the age of the decedent, the date of injury, the location of the incident, and the description of the incident, we can successfully automatically link OSHA records to the master file even without typical identifiers like decedent and establishment name. When this information is available however, our model can link these documents even more effectively. The system is likely to get production use later this summer. | Record linkage | Research | Random Forest | Python, Scikit-learn |

| Institution | Project name | Description of the project | Application | Status | Method | Software |
|---|---|---|---|---|---|---|
| U.S. Bureau of Labor Statistics | Automatic linkage of fatal injury case information to webpage articles. | To produce statistics about fatal occupational injuries in the U.S., the Census of Fatal Occupational Injuries collects and combines information from a wide variety of sources including online media such as news articles. One of the key challenges in incorporating this information is simply finding and matching it to existing case information, often in the absence of even flawed identifiers like decedent name. To address this issue this project uses machine learning, trained on previously linked documents, to automatically determine whether an automatically collected webpage article should be linked to a case already in the master file, or represents a new case that should be added. By combining a variety of noisy signals and by automatically extracting name, date, and establishment information from the article, we hope to be able to conduct this linkage automatically. Systems have already been built to automatically collect these webpages, separate the article text from the rest of the webpage, and automatically extract information like the names of people and companies mentioned in the articles, but more work remains to be done to automatically separate relevant and irrelevant articles. | Record linkage | Research | Random Forest | Python, Scikit-learn, Spacy, Apache Tika |
| U.S. Bureau of Labor Statistics | Analysis of nonresponse to the Occupational Employment Statistics (OES) Survey[65] | To gain insight into how characteristics of an establishment are associated with nonresponse, a recursive partitioning algorithm is applied to the Occupational Employment Statistics survey data to build a regression tree. The tree models an establishment's propensity to respond to the survey given certain establishment characteristics. It provides mutually exclusive cells based on the characteristics with homogeneous response propensities. This makes it easy to identify interpretable associations between the characteristic variables and an establishment's propensity to respond, something not easily done using a logistic regression propensity model. This representation is then used along with frame-level administrative wage data linked to sample data to investigate the possibility of nonresponse bias. We show that without proper adjustments the nonresponse does pose a risk of bias and is possibly nonignorable. | Nonresponse propensity modeling | Research | Regression Trees | R |

---

[65] Toth, D., Phipps, P. (2014). Regression tree models for analyzing survey response. Proceedings of the Government Statistics Section, American Statistical Association, 339–351, https://www.bls.gov/osmr/pdf/st140160.pdf (last accessed 10 December 2018).

| Institution | Project name | Description of the project | Application | Status | Method | Software |
|---|---|---|---|---|---|---|
| U.S. Bureau of Labor Statistics | Occupational Employment Statistic (OES) Occupation Autocoding | Division of Occupational of Employment Statistics (OES) uses Multinomial Logistic Regression with Stochastic Gradient Descent to develop a model to assign occupation codes to job titles received with employer survey responses. | Job title and other available information into one of many occupation codes | Development and testing | LR using SGD | Python |
| U.S. Bureau of Labor Statistics | Analysis of nonresponse to the Occupational Employment Statistics (OES) Survey[66] | Auxiliary information can increase the efficiency of survey estimators through an assisting model when the model captures some of the relationship between the auxiliary data and the study variables. Despite their superior properties, model-assisted estimators are rarely used in anything but their simplest form by statistical agencies to produce official statistics. This is due to the fact that the more complicated models that have been used in model-assisted estimation are often ill suited to the available auxiliary data. Under a model-assisted framework, we propose a regression tree estimator for a finite population total. Regression tree models are adept at handling the type of auxiliary data usually available in the sampling frame and provide a model that is easy to explain and justify. The estimator can be viewed as a post-stratification estimator where the post-strata are automatically selected by the recursive partitioning algorithm of the regression tree. We establish consistency of the regression tree estimator and a variance estimator, along with asymptotic normality of the regression tree estimator. We then compare the performance of our estimator and the coverage of the confidence intervals using our variance estimator to other survey estimators using US Bureau of Labor Statistics Occupational Employment Statistics Survey data. | Model assisted estimation | Research | Regression Trees | R (mase package) |

---

[66] McConville, K. S., Toth, D. (2017). Automated Selection of Post-Strata using a Model-Assisted Regression Tree Estimator. Cornell University arXiv:1712.05708, https://arxiv.org/pdf/1712.05708.pdf (last accessed 10 December 2018).

| Institution | Project name | Description of the project | Application | Status | Method | Software |
|---|---|---|---|---|---|---|
| U.S. Bureau of Labor Statistics | Analysis of nonresponse to the Longitudinal Occupational Employment Statistics (OES) Survey[67] | This article introduces and discusses a method for conducting an analysis of nonresponse for a longitudinal establishment survey using regression trees. The methodology consists of three parts: analysis during the frame refinement and enrollment phases, common in longitudinal surveys; analysis of the effect of time on response rates during data collection; and analysis of the potential for nonresponse bias. For all three analyses, regression tree models are used to identify establishment characteristics and subgroups of establishments that represent vulnerabilities during the data collection process. This information could be used to direct additional resources to collecting data from identified establishments in order to improve the response rate. | Nonresponse propensity modeling | Research | Regression Trees with Linear Models | R (rpms package) |
| U.S. Bureau of Labor Statistics | Outlier Detection Using Unsupervised Learning Under Informative Sampling | A Bayesian hierarchical modeling approach was developed and applied to Current Employment Statistics survey. This approach is an enhanced k-means method, and it was used to find potential outliers. | Outlier Detection | Research | k-Means | R |
| U.S. Bureau of Labor Statistics | Text Analysis of Interviewer Notes | Using text analysis and clustering (unsupervised learning) to extract information and themes from survey interviewer notes, based on data from the Consumer Expenditure Survey. We also connected the themes from interviewer notes with sample unit behavior as captured in the Contact History Instrument. | Clustering of interviewer notes | Research | Model-Based Clustering, k-Means Clustering, Bayesian Hierarchical Clustering | MATLAB and R |
| U.S. Census Bureau | Using an Autocoder to Code Industry and Occupation in the American Community Survey[68] | Every year the American Community Survey (ACS) collects industry and occupation data on nearly 2.5 million individuals. The text write-in information must then be coded, or converted to an industry or occupation numeric category code. | Multi-class classification | Test | Logistic Regression | SAS |

---

[67] Earp, M., Toth, D., Phipps, P., Oslund, C. (2018). Assessing Nonresponse in a Longitudinal Establishment Survey Using Regression Trees. Journal of Official Statistics, 34(2) 463–481, https://www.degruyter.com/downloadpdf/j/jos.2018.34.issue-2/jos-2018-0021/jos-2018-0021.pdf (last accessed 10 December 2018).
[68] Thompson, M., Kornbau, M. E., Vesely, J. (2012). Creating an Automated Industry and Occupation Coding Process for the American Community Survey. Report, https://www.census.gov/content/dam/Census/library/working-papers/2012/demo/2012-io-coding-asa-paper-final.pdf (last accessed 10 December 2018).

| Institution | Project name | Description of the project | Application | Status | Method | Software |
|---|---|---|---|---|---|---|
| U.S. Department of Agriculture NASS | Informing Sample Design for the Census of Agriculture | Random forests with boosting were applied to predict the probability of the non-respondents after mailing to respond to the Census of Agriculture. The predicted probabilities were used as one of the stratifying variables in the sample design of non-respondents. | Developing response propensity scores to inform sampling design | The response propensity scores have been used in developing a sampling plan for non-respondents. Data collection is in progress. | Random Forest with Boosting | SAS JMP |
| U.S. Department of Agriculture NASS | Informing imputation for the Census of Agriculture | Random forests with boosting are used to inform imputation of the demographics section of the Census of Agriculture | Developing response propensity scores and informing imputation | A first imputation model has been implemented, and efforts are underway to improve it. | Random Forest with Boosting | SAS and SAS JMP |
| U.S. Department of Agriculture NASS | Crop Prediction Based on the Cropland Data Layer, Administrative Data, and Survey Data | An effort to bring all available data together to enhance crop forecasts of crop yield | Blending diverse data to produce the best possible prediction of crop yield | Foundational work is being undertaken | TBD, probably Bayesian | TBD, likely R or SAS with others Source (link) |

| Institution | Project name | Description of the project | Application | Status | Method | Software |
|---|---|---|---|---|---|---|
| U.S. Department of Agriculture NASS | Machine learning for the Census of Agriculture | Machine learning methods are applied to computation needs in the production of Census of Agriculture, specifically sampling and imputation | Developing response propensity scores and informing imputation | The response propensity scores have been used in developing a sampling plan for non-respondnets. Currently, machine learning techniques are being used to inform imputation. | Random Forest with Boosting | SAS JMP |
| U.S. Department of Agriculture NASS | Questionnaire consolidation[69] | Each state had its own questionnaire version (different states were surveyed on different items at different frequencies), as it was believed that this approach reduced respondent burden. | Questionnaire consolidation | Experiment | Hierarchical Clustering | (SAS) |
| U.S. Department of Agriculture NASS | non-respondent prediction[70] | Non-response adjustment for the Census of Agriculture. Farms within each state in the US were partitioned into groups of "homogeneous response propensity" using a classification tree model. Non-response adjustments were performed within each such group based on the response rate within that group. | Classification | Experiment | Decision Tree | SAS |
| U.S. Department of Agriculture NASS | Analysis of reporting errors[71] | Prediction of respondents likely to make reporting errors based on sampling frame data. Results of this analysis could suggest reasons for the reporting errors, types of respondents to be included in questionnaire testing, and editing strategies after data collection. | Classification | Experiment | Decision Tree | SAS |

---

[69] Earp, M., Cox, S., McDaniel, J., Crouse, C. (2009). Exploring quarterly agricultural survey questionnaire version reduction scenarios. Report, https://ageconsearch.umn.edu/bitstream/234372/2/exploring%20qas%20questionnaire%20version%20reduction%20final%20print%20version.pdf (last accessed 10 December 2018).
[70] Cecere, W. (2009). 2007 Census of Agriculture Non-Response Methodology. Report.
[71] McCarthy, J. S., Earp, M. S. (2009). Who Makes Mistakes?: Using Data Mining Techniques to Analyze Reporting Errors in Total Acres Operated. Report, https://ideas.repec.org/p/ags/unasrr/234367.html (last accessed 10 December 2018).

Martin Beck; Florian Dumpert; Joerg Feuerhake: Machine Learning in Official Statistics## 5 Literature

Beck, M., Dumpert, F., Feuerhake, J. (2018). Proof of Concept Machine Learning: Abschlussbericht. Unpublished report.

Chu, K., Poirier, C. (2015). Machine Learning Documentation Initiative. UNECE Conference of European Statisticians, Workshop on the Modernisation of Statistical Production Meeting, 15–17 April 2015. https://www.unece.org/fileadmin/DAM/stats/documents/ece/ces/ge.50/2015/Topic3_Canada_paper.pdf, last accessed 28 November 2018. [Meanwhile continued by V. Todorov (UNIDO)]

Dumpert, F., Beck, M. (2017). Einsatz von Machine-Learning-Verfahren in amtlichen Unternehmensstatistiken. AStA Wirtschafts- und Sozialstatistisches Archiv, 11, 83–106.

Statistics Canada (2018). Machine Learning in Surveys Steps.

Statistisches Bundesamt (2018). Digitale Agenda des Statistischen Bundesamtes. https://www.destatis.de/DE/UeberUns/UnsereZiele/DigitaleAgenda.pdf?__blob=publicationFile, last accessed 28 November 2018.

**Referenced Literature from Chapter 4**

Alfaro, E., Gamez, M., García, N. (2013). adabag: An R Package for Classification with Boosting and Bagging. Journal of Statistical Software, 54(2).

Ballin, M., Barcaroli, G. (2013). Joint determination of optimal stratification and sample allocation using genetic algorithm. Survey Methodology, 39(2), 369–393.

Barcaroli, G., Nurra, A., Salamone, S., Scannapieco, M., Scarno, M., Summa, D. (2015). Internet as Data Source in the ISTAT Survey on ICT in Enterprises. Austrian Journal of Statistics, 44, 31–43.

Bleier, A. (2013). Practical Collapsed Stochastic Variational Inference for the HDP. Cornell University arXiv:1312.0412.

Blien, U., Hirschenauer, F., Phan thi Hong, V. (2010). Classification of regional labour markets for purposes of labour market policy. Papers in Regional Science, 89(4), 859–880.

Cagala, T. (2017). Improving data quality and closing data gaps with machine learning. IFC Bulletin.

Cecere, W. (2009). 2007 Census of Agriculture Non-Response Methodology. Report.

Ciuhu, A.-M., Caragea, N., Alexandru, C. (2017). Modeling the potential human capital on the labor market using logistic regression in R. Romanian Statistical Review, 4, 141–152.

Dallmann, A., Niebler, T., Lemmerich, F., Hotho, A. (2016, April). Extracting Semantics from Random Walks on Wikipedia: Comparing Learning and Counting Methods. In Wiki@ ICWSM.

Deneke, C., Rentzsch, R., Renard, B. Y. (2017). PaPrBaG: A machine learning approach for the detection of novel pathogens from NGS data. Scientific Reports, 7, No. 39194.63